\documentclass[%
 reprint,onecolumn,
superscriptaddress,
 amsmath,amssymb,
 aps,
]{revtex4-1}
\usepackage{color}
\usepackage{amsmath}
\usepackage{graphicx}
\usepackage{dcolumn}
\usepackage{bm}
\usepackage{comment}
\usepackage{filecontents}

\usepackage{bm}
\usepackage{amsmath,amsthm,amssymb,mathtools}
\usepackage{verbatim}
\usepackage{color}
\usepackage{tabularx}
\usepackage{xcolor}

\usepackage{hyperref}

\begin{document}

\title{\bf CMB Circular Polarization from Photon-Dark Photon Scattering}

\author{Fatemeh Masoumi Jahromi}
\affiliation{Department of Physics, Shiraz University 71454, Shiraz, Iran}

\author{Mansour Haghighat}
\affiliation{Department of Physics, Shiraz University 71454, Shiraz, Iran}

\begin{abstract}
	We investigate the generation of circular polarization ($V$-mode) in the cosmic microwave background (CMB) through forward scattering between CMB photons and a cosmological dark photon background. The standard kinetic-mixing portal yields an unobservably small signal due to stringent astrophysical, laboratory, and cosmological constraints. We therefore adopt a model-independent effective field theory with dimension-8 Euler--Heisenberg-type operators coupling the two sectors through their field strengths rather than fermion currents. Using the quantum Boltzmann formalism, we derive the evolution equations for the Stokes parameters and show that a quadrupolar asymmetry between transverse and longitudinal dark photon polarizations converts CMB linear polarization into circular polarization. We estimate the $V$-mode angular power spectrum using the scattering rate weighted by the visibility function. For a thermal relic, constrained to sub-eV masses and at most a percent of the dark matter, the benchmark sensitivity $\ell(\ell+1)C_\ell^{VV}/(2\pi)\gtrsim0.01\,\mathrm{nK}^2$ requires $\Lambda/(C_1-C_2)^{1/4}\lesssim1.8\,\mathrm{MeV}$, where $\Lambda$ is the effective scale and $C_{1,2}$ are Wilson coefficients. A cold population produced by heavy particle decay yields a mass-independent signal, reaching the same benchmark for $\Lambda/(C_1-C_2)^{1/4}\lesssim6.5\,\mathrm{MeV}$, or about $27\,\mathrm{MeV}$ for a colder and more abundant population. These scales lie within an order of magnitude of the bound suggested by stellar energy loss. Our results establish CMB circular polarization as a complementary probe of higher-dimensional photon-dark photon interactions, sensitive to the mass, abundance, and thermal state of the dark photon background.
\end{abstract}

\maketitle

\section{Introduction}
The polarization of the cosmic microwave background (CMB) is a cornerstone
observable of modern cosmology \cite{hu2002cmb}. Its linear component, described by the
Stokes parameters $Q$ and $U$ or equivalently by the $E$- and $B$-mode
decomposition, arises from Thomson scattering at recombination, where local
quadrupole anisotropies in the photon distribution are converted into linear
polarization
\cite{1996AnPhy.246...49K,chandrasekhar_radiative_1960,seljak1996line,zaldarriaga1997allsky,kamionkowski1997statistics,hu1997primer}. This signal has
been measured with high precision by WMAP \cite{bennett2013wmap9}, Planck
\cite{planck2018params,planck2018spectra}, BICEP/Keck \cite{bicepkeck2021}, and the
Atacama Cosmology Telescope \cite{act2020dr4spectra}. The lensing $B$-mode is now detected at
high significance as well \cite{act2023dr6lensing}. CMB linear polarization is therefore a well-established
and precisely tested prediction of the $\Lambda$CDM model. However, the $V$-mode, which describes circular polarization,
has received almost no attention, because Thomson scattering cannot
generate circular polarization, and in
standard electrodynamics $V$ vanishes \cite{1996AnPhy.246...49K,chandrasekhar_radiative_1960}. Therefore, any detection of $V$-mode would be a signal of physics beyond
$\Lambda$CDM. Consequently, a number of experiments have searched for a nonzero $V$
signal, with upper limits placed by experiments such as MIPOL \cite{mainini2013improved}
and SPIDER \cite{2017ApJ...844..151N}. The most recent and most stringent limit comes from the
CLASS experiment, which has placed upper limits of $\ell(\ell+1)C_\ell^{VV}/(2\pi)$ in
the range $0.4$--$13.5\,\mu\mathrm{K}^2$ over angular scales $1\le\ell\le120$
\cite{2020ApJ...889..105P}, improving on earlier bounds by two orders of magnitude.

These improving experimental sensitivities motivate us to identify particle-physics mechanisms capable of sourcing a nonzero $V$-mode. Several
such mechanisms have already been proposed in the literature (see, e.g.,
\cite{alexander2020physics} for a review and a general Boltzmann-hierarchy formalism),
including circular
polarization generated in the presence of a background magnetic field
\cite{giovannini2010circular,giovannini2009vmode}, from
Lorentz-noninvariant operators in the QED sector, or from spacetime noncommutativity
\cite{zarei2010generation,tizchang2016noncommutative}; from a Chern--Simons-type photon coupling
\cite{2009PhRvD..79f3524A}; from Faraday conversion in primordial magnetic fields
\cite{cooray2003circularly}; and from conventional sources of cosmic birefringence
\cite{montero2018exploring}. Other proposed sources include axion--photon mixing
\cite{agarwal2008signatures,finelli2009rotation} and interactions between CMB photons
and cosmic neutrinos or other fermions \cite{mohammadi2014evidence,bartolo2019cmb,haghighat2020sterile}.
The forward
scattering between ordinary CMB photons themselves, mediated by the Euler-Heisenberg
interaction, has likewise been shown to source a $V$-mode via Faraday conversion
\cite{hoseinpour2020cmb,sawyer2015photonphoton}, also in the presence of primordial vector
and tensor perturbations \cite{inomata2019circular}.

In this paper we consider the scattering between ordinary CMB photons and a
cosmological background of dark photons, $\gamma+\gamma'\to\gamma+\gamma'$, as the
underlying mechanism for generating $V$. Dark photons are the gauge bosons of a hidden
$U(1)'$ gauge group and are among the best motivated portals between the visible and dark
sectors \cite{1986PhLB..166..196H} (for a review, see \cite{fabbrichesi2020dark}). Depending on their mass, abundance, and
production history, dark photons can behave as dark radiation, warm matter, or
cold dark matter, and their parameter space is constrained by an extensive
program of laboratory searches, stellar and supernova cooling, and cosmological
observations \cite{caputo2021dark,caputo2025heavy}.

The standard way of connecting the two sectors is the renormalizable kinetic-mixing
operator $-(\epsilon/2)F_{\mu\nu}F^{\prime\mu\nu}$, which after diagonalization gives the
dark photon a direct coupling $\epsilon e A'_\mu J^\mu_{\rm EM}$ to the electromagnetic
current. It is precisely this tree-level coupling to charged matter that makes the
minimal scenario so strongly constrained. Stellar and supernova cooling, beam-dump and
fixed-target searches, and cosmological bounds on $\Delta N_{\rm eff}$ together require
$\epsilon\lesssim10^{-14}$ for sub-keV dark photons, and $\epsilon\lesssim10^{-8}$--$10^{-4}$
in the MeV--GeV range \cite{fabbrichesi2020dark,caputo2021dark,caputo2025heavy}. These
bounds have a decisive consequence for CMB circular polarization. If the
$\gamma\gamma'$ interaction is the one induced by kinetic mixing through the electron
loop, the coefficient of the resulting dimension-8 operator is fixed to
$C/\Lambda^4\sim\epsilon^2\alpha^2/m_e^4$, that is
$\Lambda\sim m_e/(\epsilon\alpha)^{1/2}\gtrsim\mathcal{O}(10\ \mathrm{TeV})$ for
$\epsilon\lesssim10^{-14}$. Since, as we show below, the $V$-mode source is proportional
to $\Lambda^{-4}$, at such scales the dimensionless parameter
$\beta\equiv\dot\tau_{\gamma\gamma'}/\dot\tau_{e\gamma}$ of Eq.~(\ref{eq:ratio}) is
suppressed by more than twenty orders of magnitude relative to the values required for
observability. The minimal kinetic-mixing portal therefore cannot yield a detectable CMB
$V$-mode, and a different interaction structure is needed.

We consequently adopt a model-independent effective field theory (EFT) description in
which the two sectors communicate through higher-dimensional operators alone, with no
renormalizable mixing term and no tree-level coupling of the dark photon to Standard
Model fermions. The leading gauge-invariant, Lorentz-invariant interaction between the
two field strengths appears at dimension 8, in direct analogy with the nonlinear photon
self-interactions induced by the electron loop in ordinary QED
\cite{2006physics...5038H}. The crucial difference with respect to kinetic mixing is that
this operator couples only to field strengths, not to fermion currents, so that stellar
production proceeds through photon fusion $\gamma\gamma\to\gamma'\gamma'$ rather than
plasmon decay or bremsstrahlung, and is parametrically much more suppressed. The scale
$\Lambda$ and the Wilson coefficients $C_1$, $C_2$ are then independent parameters of the
low-energy theory \cite{georgi1993effective}, to be fixed either by a specified
ultraviolet completion or, as we do here, by observation. As an illustrative example,
a heavy fermion $\Psi$ carrying a millicharge $\epsilon_\Psi e$ under
electromagnetism and a dark charge $e_D$ under $U(1)'$ generates the operator at one loop
with $C/\Lambda^4\sim\epsilon_\Psi^2\alpha_D^2/M_\Psi^4$; here $\epsilon_\Psi$ is a charge
ratio of the mediator and is unrelated to the kinetic-mixing parameter $\epsilon$
constrained above. However, this is only one possible ultraviolet completion among many.
Our EFT framework is entirely agnostic to the high-energy details and captures the
low-energy phenomenology of any underlying theory that yields photon--dark-photon
scattering without inducing excluded tree-level fermion currents. We do not commit to
any such completion, and treat Eq.~(\ref{eq:lint}) as the object to be constrained.

Using the quantum Boltzmann equation for the photon density matrix, we derive the
evolution equations for the Stokes parameters and show that circular polarization is
sourced only when the CMB carries linear polarization and the dark-photon
polarization distribution possesses a quadrupolar anisotropy. The amplitude of the effect is governed by the
dimensionless ratio $\beta$, which scales as
\begin{equation}
	\beta\;\propto\;\frac{C_1-C_2}{\Lambda^4}\;
	\frac{m_{\gamma'}^2\,n_{\gamma'}}{T_{\gamma'}}\;,
	\label{eq:beta-intro}
\end{equation}
so that the interaction strength enters only through $\Lambda$ and the Wilson
coefficients, while all information about the production history of the dark-photon
background is contained in the phase-space combination
$m_{\gamma'}^2 n_{\gamma'}/T_{\gamma'}$. This separation is what makes the EFT treatment
powerful. The same expression for $\beta$ applies to any production mechanism, and the
observable depends on $\Lambda$ and on the dark-photon phase-space density
independently.

Finally, we stress that dedicated astrophysical and laboratory constraints on the
operator in Eq.~(\ref{eq:lint}) have not been computed in the literature, since existing
bounds are derived for kinetic mixing or for axion-like couplings and do not apply
directly to a pure field-strength contact interaction. An order-of-magnitude
dimensional-analysis estimate of the stellar energy-loss rate from
$\gamma\gamma\to\gamma'\gamma'$ in horizontal-branch stars gives
$\Lambda\gtrsim\mathcal{O}(10\ \mathrm{MeV})$ for Wilson coefficients of order unity.
The scales required for an observable $V$-mode, derived in Sec.~\ref{V}, lie within
an order of magnitude of this estimate and can exceed it for a cold and abundant
decay-produced population. We leave the calculation of the astrophysical constraints on
this operator to future work.
Independently of this, our analysis establishes CMB circular
polarization as a new probe of this class of interactions. Whereas laboratory searches
constrain $\Lambda$ through high-energy scattering, the $V$-mode signal depends on the
combination of $\Lambda$ with the phase-space density of the dark-photon background, and
is therefore sensitive to its production history and thermal state in a way that no
terrestrial experiment is.

The paper is organized as follows. In Sec.~\ref{II} we review the Stokes parameters and
the quantum Boltzmann equation. In Sec.~\ref{III} we introduce the effective
photon--dark-photon interaction and derive the evolution equations for the polarization.
In Sec.~\ref{IV} we estimate the $V$-mode power spectrum, and in Sec.~\ref{V} we evaluate
it for thermal and non-thermal dark-photon populations. We conclude in Sec.~\ref{VI}.

\section{Polarization Description}\label{II}
 To analyze the CMB polarization we begin by introducing the Stokes parameters, and the equation of motion for the evolution of polarization \cite{1996AnPhy.246...49K,kosowsky2001cmb}. For an electromagnetic wave $ \vec E $ propagating along the $\hat z$ direction,
 polarization properties can be described by means of the Stokes parameters \emph{I}, \emph{Q}, \emph{U} and \emph{V}. In the polarization basis, they are defined as the following time-averaged quantities
\begin{align}\label{Stokesparameters}
	I &\equiv \langle E_{x}^{2}\rangle + \langle E_{y}^{2}\rangle \ , &
	Q &\equiv \langle E_{x}^{2}\rangle - \langle E_{y}^{2}\rangle \ , \\
	U &\equiv \langle 2 E_{x} E_{y} \cos(\phi_{x}-\phi_{y}) \rangle \ , &
	V &\equiv \langle 2 E_{x} E_{y} \sin(\phi_{x}-\phi_{y}) \rangle \ .
\end{align}
where $ E_{x} $ and $ E_{y} $ are the amplitudes, and $ \phi_{x} $ and $ \phi_{y} $ are the phases of the wave in the x-y plane. The \emph{I} parameter represents the total intensity of radiation. For unpolarized radiation $ Q=U=V=0 $, while linearly polarized radiation is indicated by non-zero $ Q $ and/or $ U $, which are coordinate dependent parameters. The Stokes-V parameter quantifies the circular polarization and like $ I $ does not depend on the chosen coordinate system. The evolution of CMB polarization is described by the Boltzmann equation, which gives the evolution of the phase space distribution function $ f $
\begin{equation}
\frac{df}{dt}=C[f],
\end{equation}
 In quantum mechanics, in the polarization state space the intensity and polarization distribution of an ensemble of photons is encoded in the density matrix. In the linear polarization basis the density matrix in terms of the classical Stokes parameters is characterized by
\begin{align}\label{e14}
	\rho =\frac{1}{2}\begin{pmatrix}I+Q & U-iV \\
		U+iV & I-Q \\
	\end{pmatrix},
\end{align} 

where the density matrix elements $\rho_{ij}$ are related to the number operator $ \mathcal{\hat {D}}_{ij}(\mathbf{k})=\hat a_{i}^{\dag}(\mathbf{k})\hat a_{j}(\mathbf{k})$  as follows

\begin{align}\label{e16}
\langle \mathcal{\hat {D}}_{ij}(\mathbf{k})\rangle=(2\pi)^{3}2k^{0}\delta^{(3)}(0)\rho_{ij}(\mathbf{k}).
\end{align}
where $ k^{0} $ is the energy of photons. Now the Boltzmann equation in terms of $ \rho $ is given by
\begin{align}
	(2\pi)^{3}2k^{0}\delta^{(3)}(0)\frac{d}{dt}\rho_{ij}(\mathbf{k})=i\langle[\mathcal{\hat H}_{int} (t),\mathcal{\hat {D}}_{ij}(\mathbf{k})]\rangle-\frac{1}{2}\int^{+\infty}_{-\infty} dt \langle[\mathcal{\hat H}_{int} (t),[\mathcal{\hat H}_{int} (0),\mathcal{\hat {D}}_{ij}(\mathbf{k})]]\rangle
	\label{BE}
\end{align}
where $\mathcal{\hat H}_{int}$ is the interacting Hamiltonian. The first term on the right-hand side is referred as the forward-scattering term while the second one is the usual collision term. In the case of linear Maxwellian electrodynamics, the time evolution of the Stokes parameter \emph{V} is equal to zero, and no circular polarization occurs.

\section{Photon-dark photon scattering and CMB Polarization}\label{III}
To derive the photon--dark photon interaction term, we employ the Euler--Heisenberg
effective Lagrangian \cite{2006physics...5038H}. This effective theory, valid at energies lower than the electron
mass, extends classical Maxwell electrodynamics via quantum corrections, thereby
enabling the self-interaction of gauge bosons. The Euler-Heisenberg effective Lagrangian
is given by
\begin{equation}
	\mathcal{L}_{\rm eff} =
	\frac{\alpha^2}{90m_e^4}
	\left[ (F_{\mu\nu}F^{\mu\nu})^2 + \frac{7}{4}(F_{\mu\nu}\tilde{F}^{\mu\nu})^2 \right],
	\label{eq:EH}
\end{equation}
where $\alpha$ is the fine structure constant and $ m_e $ is the electron mass. $\tilde{F}^{\mu\nu} = \frac{1}{2}\epsilon^{\mu\nu\alpha\beta}F_{\alpha\beta}$
is the dual field strength tensor.

We now extend this framework to describe the scattering $\gamma + \gamma' \to \gamma + \gamma'$,
where $\gamma'$ denotes the dark photon with field strength $F'_{\mu\nu}$.
By analogy with the Euler--Heisenberg construction, the leading interaction between
the two sectors arises at dimension 8. Replacing one pair of SM field strengths in
Eq.~\eqref{eq:EH} with dark photon field strengths, the effective interaction Lagrangian
responsible for $\gamma\gamma' \to \gamma\gamma'$ scattering takes the form
\begin{equation}
	\mathcal{L}_{\rm int} =
	\frac{1}{\Lambda^4}
	\left[
	C_{1}\,(F_{\mu\nu}F^{\prime\mu\nu})^2
	+ C_{2}\,(F_{\mu\nu}\tilde{F}^{\prime\mu\nu})^2
	\right],
	\label{eq:lint}
\end{equation}
where $\Lambda$ is the effective scale of the interaction, and $C_1$, $C_2$ are dimensionless
Wilson coefficients. As with the prefactor $\alpha^2/90m_e^4$ in Eq.~\eqref{eq:EH}, which is
fixed by the electron loop that generates it, $C_1$ and $C_2$ would be calculable once the
particle content mediating the coupling between the two $U(1)$ sectors, i.e. what runs
in the loop, is specified \cite{georgi1993effective}. Since we work with a general
Euler--Heisenberg-type effective theory without committing to a particular UV completion,
we leave $C_1$ and $C_2$ undetermined and carry their combination $(C_1-C_2)$ explicitly
through our final results, rather than fixing it to a specific value.
The two independent operators in Eq.~\eqref{eq:lint} are the complete set of
Lorentz-invariant, parity-even, dimension-8 structures built from one SM and
one dark field strength, and are therefore the lowest-order operators that can
mediate the process $\gamma + \gamma' \to \gamma + \gamma'$.
By dimensional analysis, the resulting scattering amplitude scales as
$\mathcal{M} \sim s^2/\Lambda^4$, and the total cross section as
$\sigma \sim s^3/\Lambda^8$, so that the effective theory description is valid for
centre-of-mass energies $\sqrt{s} \ll \Lambda$.

The free photon and dark photon fields in terms of creation and annihilation
operators in the Coulomb (radiation) gauge can be written as
\begin{equation}
	A^\mu(x) = \int \frac{d^3\mathbf{q}}{(2\pi)^3 2q^0}
	\left[ \hat{a}_r(q)\,\epsilon_r^\mu(q)\,e^{-iq\cdot x}
	+ \hat{a}^\dagger_r(q)\,\epsilon_r^{*\mu}(q)\,e^{iq\cdot x} \right],
	\label{eq:photon_field}
\end{equation}
\begin{equation}
	A^{\prime\mu}(x) = \int \frac{d^3\mathbf{q'}}{(2\pi)^3 2q'^0}
	\left[ \hat{b}_{r'}(q')\,\epsilon_{r'}^\mu(q')\,e^{-iq'\cdot x}
	+ \hat{b}^\dagger_{r'}(q')\,\epsilon_{r'}^{*\mu}(q')\,e^{iq'\cdot x} \right],
	\label{eq:darkphoton_field}
\end{equation}
with $\epsilon_r^\mu(q) = (0,\boldsymbol{\epsilon}_r(q))$, $r = 1,2$, denoting the two transverse
polarization four-vectors of the (massless) photon, and $\epsilon_{r'}^\mu(q')$,
$r' = 1,2,3$, the two transverse and one longitudinal polarization four-vectors of the
(massive) dark photon, satisfying $q\cdot\epsilon_r(q)=0$ and
$q'\cdot\epsilon_{r'}(q')=0$; $q$ and $q'$ stand for the corresponding four-momenta, with
$q^0 = |\mathbf{q}|$ for the photon and $q'^0=\sqrt{|\mathbf{q}'|^2+m_{\gamma'}^2}$ for the
dark photon. The creation and annihilation
operators of photons and dark photons satisfy the canonical commutation relation as
\begin{equation}
	[\hat{a}_i(q),\hat{a}^\dagger_j(p)] = (2\pi)^3 2q^0\,\delta_{ij}\delta^{(3)}(\mathbf{q}-\mathbf{p}),
	\qquad
	[\hat{b}_i(q'),\hat{b}^\dagger_j(p')] = (2\pi)^3 2q'^0\,\delta_{ij}\delta^{(3)}(\mathbf{q'}-\mathbf{p'}).
	\label{eq:commutation}
\end{equation}
From (\ref{eq:lint}) and $H_{\rm int} = \int d^3x\,\mathcal{H}_{\rm int}(x)$, the interaction
Hamiltonian describing $\gamma(q)+\gamma'(q')\to\gamma(p)+\gamma'(p')$ is given by
\begin{align}
	H_{\rm int}(t) &= 4
	\int d\mathbf{q}\,d\mathbf{q'}\,d\mathbf{p}\,d\mathbf{p'}\;(2\pi)^3\delta^3(\mathbf{q}+\mathbf{q'}-\mathbf{p}-\mathbf{p'})\,
	\mathcal{M}(\mathbf{q},r;\,\mathbf{q'},r';\,\mathbf{p},s;\,\mathbf{p'},s')\,
	e^{-i(q^0+q'^0-p^0-p'^0)t} \notag\\
	&\quad \times a^\dagger_s(p)\,a_r(q)\,b^\dagger_{s'}(p')\,b_{r'}(q'),
	\label{eq:hint}
\end{align}
where $d\mathbf{q}\equiv d^3\mathbf{q}/[(2\pi)^3\,2q^0]$ is the Lorentz-invariant phase-space measure
(and likewise $d\mathbf{q'}$, $d\mathbf{p}$, $d\mathbf{p'}$ for the other momenta), and the amplitude $\mathcal{M}$
is as follows
\begin{align}
	\mathcal{M}(\mathbf{q},r;\,\mathbf{q'},r';\,\mathbf{p},s;\,\mathbf{p'},s') &=
	\frac{2}{\Lambda^4}\left(C_{1}\, g^{\mu\alpha}g^{\nu\beta}g^{\rho\gamma}g^{\sigma\delta}
	+\frac{C_{2}}{4} \,\epsilon^{\mu\nu\alpha\beta}\epsilon^{\rho\sigma\gamma\delta} \right)
	\notag\\
	&\times
	\Big[
	(q_\mu\epsilon_{r\nu}(q)-q_\nu\epsilon_{r\mu}(q))
	(q'_\alpha\epsilon_{r'\beta}(q')-q'_\beta\epsilon_{r'\alpha}(q'))
	(p_\rho\epsilon_{s\sigma}(p)-p_\sigma\epsilon_{s\rho}(p))
	(p'_\gamma\epsilon_{s'\delta}(p')-p'_\delta\epsilon_{s'\gamma}(p'))
	\notag\\
	&+
	(q_\mu\epsilon_{r\nu}(q)-q_\nu\epsilon_{r\mu}(q))
	(p'_\alpha\epsilon_{s'\beta}(p')-p'_\beta\epsilon_{s'\alpha}(p'))
	(q'_\gamma\epsilon_{r'\delta}(q')-q'_\delta\epsilon_{r'\gamma}(q'))
	(p_\rho\epsilon_{s\sigma}(p)-p_\sigma\epsilon_{s\rho}(p))
	\Big].
	\label{eq:amplitude}
\end{align}

Using the expectation value relations between the photon and dark photon creation and
annihilation operators
\begin{align}
	\langle a_1 a_2\cdots b_1 b_2\cdots\rangle
	&= \langle a_1 a_2\cdots\rangle\langle b_1 b_2\cdots\rangle, \notag\\
	\langle \hat{a}^\dagger_i(q)\hat{a}_j(p)\rangle
	&= 2p^0(2\pi)^3\delta^3(\mathbf{q}-\mathbf{p})\,\rho_{ij}(\mathbf{p}), \notag\\
	\langle \hat{b}^\dagger_i(q')\hat{b}_j(p')\rangle
	&= 2p'^0(2\pi)^3\delta^3(\mathbf{q'}-\mathbf{p'})\,\rho_{ij}(\mathbf{p'}),
	\label{eq:expectations}
\end{align}
and substituting (\ref{eq:hint}) into (\ref{BE}), the evolution equation
for the density matrix elements of CMB photons considering the forward scattering term
in the Boltzmann equation reads
\begin{align}
	\dot{\rho}_{ij}(\mathbf{x},\mathbf{k}) &= \frac{4i}{\Lambda^4\,k^0}
	\int\frac{d^3\mathbf{q'}}{(2\pi)^3 2q'^0}
	\sum_{s',r'=1}^{3}\rho_{s'r'}(\mathbf{x},\mathbf{q'})
	\notag\\
	&\quad\times
	(\delta_{ir}\rho_{sj}(\mathbf{x},\mathbf{k}) - \delta_{js}\rho_{ir}(\mathbf{x},\mathbf{k}))
	\left(C_{1}g^{\mu\alpha}g^{\nu\beta}g^{\rho\gamma}g^{\sigma\delta}
	+\frac{C_{2}}{4}\epsilon^{\mu\nu\alpha\beta}\epsilon^{\rho\sigma\gamma\delta}\right)
	\notag\\
	&\quad\times
	\Big[
	(k_\mu\epsilon_{r\nu}(k)-k_\nu\epsilon_{r\mu}(k))
	(q'_\alpha\epsilon_{r'\beta}(q')-q'_\beta\epsilon_{r'\alpha}(q'))
	(k_\rho\epsilon_{s\sigma}(k)-k_\sigma\epsilon_{s\rho}(k))
	(q'_\gamma\epsilon_{s'\delta}(q')-q'_\delta\epsilon_{s'\gamma}(q'))
	\notag\\
	&\quad+
	(k_\mu\epsilon_{r\nu}(k)-k_\nu\epsilon_{r\mu}(k))
	(q'_\alpha\epsilon_{s'\beta}(q')-q'_\beta\epsilon_{s'\alpha}(q'))
	(q'_\gamma\epsilon_{r'\delta}(q')-q'_\delta\epsilon_{r'\gamma}(q'))
	(k_\rho\epsilon_{s\sigma}(k)-k_\sigma\epsilon_{s\rho}(k))
	\Big].
	\label{eq:rho_evolution}
\end{align}

To evaluate (\ref{eq:rho_evolution}) more easily we align the line of sight $\hat{n}$
with the $z$ axis, and fix the frame as $k \parallel \hat{z}$ and write the photon and
dark photon momentum and polarization four-vectors in the following form
\begin{align}
	k^\mu &= (k^0,\,0,\,0,\,k^0), \notag\\
	\epsilon_1(k) &= (0,\,1,\,0,\,0), \notag\\
	\epsilon_2(k) &= (0,\,0,\,1,\,0), \notag\\
	q^{\prime\mu} &= (q'^0,\,q'\sin\theta\cos\phi,\,q'\sin\theta\sin\phi,\,q'\cos\theta), \notag\\
	\epsilon_1(q') &= (0,\,\cos\theta\cos\phi,\,\cos\theta\sin\phi,\,-\sin\theta), \notag\\
	\epsilon_2(q') &= (0,\,-\sin\phi,\,\cos\phi,\,0), \notag\\
	\epsilon_3(q') &= \left(\frac{q'}{m_{\gamma'}},\,\frac{q'^0}{m_{\gamma'}}\sin\theta\cos\phi,\,
	\frac{q'^0}{m_{\gamma'}}\sin\theta\sin\phi,\,
	\frac{q'^0}{m_{\gamma'}}\cos\theta\right).
	\label{eq:frame}
\end{align}

We can expand the dark photon distribution function about a uniform unpolarized
distribution as
\begin{equation}
	\rho_{s'r'}(\mathbf{x},t,q',\hat{q}') = \rho^{(0)}_{s'r'}(t,q') + \rho^{(1)}_{s'r'}(\mathbf{x},t,q',\hat{q}')
	\label{eq:dp_expansion}
\end{equation}
where $\rho^{(0)}_{s'r'} = I_0(q')/3$ for $s'=r'$ and $\rho^{(0)}_{s'r'} = 0$ for
$s'\neq r'$. It is easy to verify that (\ref{eq:rho_evolution}) vanishes when evaluated
in terms of $\rho^{(0)}$, which means that the distribution of dark photons needs to be
anisotropic to contribute to the polarization of CMB. So we turn to the first order
perturbation evaluating (\ref{eq:rho_evolution}) and consider the first order density
matrix of dark photons as the following
\begin{equation}
	\rho^{(1)}_{s'r'} = \frac{1}{3}\left(q'^0\frac{\partial I_0(q')}{\partial q'^0}\right)
	\begin{pmatrix}
		I(\mathbf{x},\hat{q}') + \delta(\mathbf{x},\hat{q}') & 0 & 0 \\
		0 & I(\mathbf{x},\hat{q}') + \delta(\mathbf{x},\hat{q}') & 0 \\
		0 & 0 & I(\mathbf{x},\hat{q}') - 2\delta(\mathbf{x},\hat{q}')
	\end{pmatrix}
	\label{eq:rho1}
\end{equation}
where $\delta$ indicates the difference between the intensity of the transverse and the
longitudinal modes, and $I_0(q') = \exp\!\left(-\frac{q'^0-\mu}{T_{\gamma'}}\right)$
is the Maxwell-Boltzmann distribution function describing the unperturbed distribution of
dark photons. In the non-relativistic limit $q'^0 \approx m_{\gamma'}$, keeping only the
leading term consistent with the order at which Eq.~(\ref{eq:rho_evolution}) was derived,
using (\ref{eq:rho_evolution}) the time evolution of the Stokes parameters is given by
\begin{align}
	\dot{I}(k) &= 0,
	\label{eq:Idot} \\
	\dot{Q}(k) &= -\frac{4(C_{1}-C_{2})}{\pi \Lambda^4}
	\frac{m^2_{\gamma'} n_{\gamma'} k^0}{ T_{\gamma'}}
	\int d^2\hat{q}'\,(1-\cos^2\theta)\sin 2\phi\;\delta(\mathbf{x},\hat{q}')\,V(k),
	\label{eq:Qdot} \\
	\dot{U}(k) &= \frac{4(C_{1}-C_{2})}{\pi \Lambda^4}
	\frac{m^2_{\gamma'} n_{\gamma'} k^0}{ T_{\gamma'}}
	\int d^2\hat{q}'\,(1-\cos^2\theta)\cos 2\phi\;\delta(\mathbf{x},\hat{q}')\,V(k),
	\label{eq:Udot} \\
	\dot{V}(k) &= \frac{4(C_{1}-C_{2})}{\pi \Lambda^4}
	\frac{m^2_{\gamma'} n_{\gamma'} k^0}{ T_{\gamma'}}
	\int d^2\hat{q}'\,(1-\cos^2\theta)
	\big[\sin 2\phi\;Q(k)-\cos 2\phi\;U(k) \big]\,\delta(\mathbf{x},\hat{q}'),
	\label{eq:Vdot}
\end{align}
where $\dot{I}(k)=0$ since forward scattering involves no exchange of energy or momentum
between the colliding beams, leaving the photon intensity unaffected, and
$n_{\gamma'} = \int\frac{d^3q'}{(2\pi)^3}I_0(q')$ is the number density of dark
photons. On the right hand side we can see that linear polarization appears as the source
for generating circular polarization, and vice versa. The factor $m^2_{\gamma'}$ reflects
the role of the longitudinal polarization. Since the longitudinal component of $F'_{\mu\nu}$ is
proportional to $m_{\gamma'}$, the asymmetry $\delta$ between transverse and
longitudinal occupations couples to the photons only for a massive dark photon, and the
source of circular polarization vanishes in the relativistic limit $m_{\gamma'}\ll q'^0$,
as we have verified by repeating the calculation in that limit.
\section{Power Spectrum of CMB Circular Polarization}\label{IV}
In this section we estimate the CMB circular polarization power spectrum due to the
interaction with the dark photons. Equations~(\ref{eq:Qdot})--(\ref{eq:Vdot}) are
expressed in terms of proper time $t$ and the physical photon energy $k^0$; to substitute
them as source terms in the Boltzmann equations below, we convert to conformal time
$\eta$, with $d/dt = a(\eta)^{-1}d/d\eta$, and to the comoving photon energy
$k_c^0\equiv a(\eta)k^0$. Expanding $\delta(\mathbf{x},\hat{q}')$ in terms of spherical harmonics as
$\delta(\mathbf{x},\hat{q}') = \sum_{\ell m}\delta_{\ell m}(\mathbf{x})Y_{\ell m}(\theta,\phi)$ and writing
the evolution equations~(\ref{eq:Qdot})--(\ref{eq:Vdot}) in terms of the Stokes brightness
perturbations, we obtain
\begin{align}
	\frac{d}{d\eta}\Delta_Q(\mathbf{x},k_c) &= \frac{16}{\pi}\sqrt{\frac{2\pi}{15}}
	\frac{m^2_{\gamma'} n_{\gamma'} k^0_c(C_{1}-C_{2})}{\Lambda^4 T_{\gamma'}}\,
	\mathrm{Im}(\delta_{2,2}(\mathbf{x}))\,\Delta_V(\mathbf{x},k_c),
	\label{eq:Qdot_fourier} \\
	\frac{d}{d\eta}\Delta_U(\mathbf{x},k_c) &= \frac{16}{\pi}\sqrt{\frac{2\pi}{15}}
	\frac{m^2_{\gamma'} n_{\gamma'} k^0_c(C_{1}-C_{2})}{\Lambda^4 T_{\gamma'}}\,
	\mathrm{Re}(\delta_{2,2}(\mathbf{x}))\,\Delta_V(\mathbf{x},k_c),
	\label{eq:Udot_fourier} \\
	\frac{d}{d\eta}\Delta_V(\mathbf{x},k_c) &= -\frac{16}{\pi}\sqrt{\frac{2\pi}{15}}
	\frac{m^2_{\gamma'} n_{\gamma'} k^0_c(C_{1}-C_{2})}{\Lambda^4 T_{\gamma'}}
	\Big[
	\mathrm{Im}(\delta_{2,2}(\mathbf{x}))\,\Delta_Q(\mathbf{x},k_c)
	+ \mathrm{Re}(\delta_{2,2}(\mathbf{x}))\,\Delta_U(\mathbf{x},k_c)
	\Big],
	\label{eq:Vdot_fourier}
\end{align}
where $\delta_{2,2}(\mathbf{x})$ is the local quadrupole of $\delta(\mathbf{x},\hat{q}')$,
the asymmetry between the intensities of the transverse and longitudinal modes of the
dark photons. The source term therefore turns out to be set by the quadrupole moment of
the dark-photon polarization asymmetry.

It is convenient to combine the linear-polarization brightness perturbations into
\begin{equation}
	\Delta_P^\pm(\mathbf{x},k_c) \equiv \Delta_Q(\mathbf{x},k_c)\pm i\Delta_U(\mathbf{x},k_c),
	\label{eq:DeltaPpm}
\end{equation}
the standard combination used to project onto $E$- and $B$-modes. Adding and subtracting
Eqs.~(\ref{eq:Qdot_fourier}) and (\ref{eq:Udot_fourier}) gives
\begin{align}
	\frac{d}{d\eta}\Delta_P^+(\mathbf{x},k_c) &= i\,\frac{16}{\pi}\sqrt{\frac{2\pi}{15}}
	\frac{m^2_{\gamma'}n_{\gamma'}k^0_c(C_1-C_2)}{\Lambda^4T_{\gamma'}}\,
	\delta^*_{2,2}(\mathbf{x})\,\Delta_V(\mathbf{x},k_c),
	\label{eq:Ppdot} \\
	\frac{d}{d\eta}\Delta_P^-(\mathbf{x},k_c) &= -i\,\frac{16}{\pi}\sqrt{\frac{2\pi}{15}}
	\frac{m^2_{\gamma'}n_{\gamma'}k^0_c(C_1-C_2)}{\Lambda^4T_{\gamma'}}\,
	\delta_{2,2}(\mathbf{x})\,\Delta_V(\mathbf{x},k_c).
	\label{eq:Pmdot}
\end{align}
Fourier transforming Eqs.~(\ref{eq:Ppdot}), (\ref{eq:Pmdot}), and (\ref{eq:Vdot_fourier})
turns the real-space product of $\delta_{2,2}(\mathbf{x})$ with $\Delta_Q$, $\Delta_U$, or
$\Delta_V$ into a convolution,
\begin{align}
	\frac{d}{d\eta}\Delta_P^+(\mathbf{K},k_c) &= i\,\frac{16}{\pi}\sqrt{\frac{2\pi}{15}}
	\frac{m^2_{\gamma'}n_{\gamma'}k^0_c(C_1-C_2)}{\Lambda^4T_{\gamma'}}
	\int\frac{d^3\mathbf{K}_1}{(2\pi)^3}\,\delta^*_{2,2}(\mathbf{K}-\mathbf{K}_1)\,
	\Delta_V(\mathbf{K}_1,k_c),
	\label{eq:Ppdot_fourier} \\
	\frac{d}{d\eta}\Delta_P^-(\mathbf{K},k_c) &= -i\,\frac{16}{\pi}\sqrt{\frac{2\pi}{15}}
	\frac{m^2_{\gamma'}n_{\gamma'}k^0_c(C_1-C_2)}{\Lambda^4T_{\gamma'}}
	\int\frac{d^3\mathbf{K}_1}{(2\pi)^3}\,\delta_{2,2}(\mathbf{K}-\mathbf{K}_1)\,
	\Delta_V(\mathbf{K}_1,k_c),
	\label{eq:Pmdot_fourier} \\
	\frac{d}{d\eta}\Delta_V(\mathbf{K},k_c) &= -\frac{16}{\pi}\sqrt{\frac{2\pi}{15}}
	\frac{m^2_{\gamma'}n_{\gamma'}k^0_c(C_1-C_2)}{\Lambda^4T_{\gamma'}}
	\int\frac{d^3\mathbf{K}_1}{(2\pi)^3}\Big[
	\mathrm{Im}(\delta_{2,2}(\mathbf{K}-\mathbf{K}_1))\,\Delta_Q(\mathbf{K}_1,k_c) \notag\\
	&\qquad\qquad\qquad\qquad\qquad\qquad\qquad\quad
	+\mathrm{Re}(\delta_{2,2}(\mathbf{K}-\mathbf{K}_1))\,\Delta_U(\mathbf{K}_1,k_c)\Big].
	\label{eq:Vdot_K}
\end{align}

Considering the FRW spacetime as the background spacetime, and working in the frame with
$\mathbf{K}\parallel\hat{z}$ so that $\mu\equiv\hat{n}\cdot\hat{K}$ is the cosine between the
photon direction and the wavevector, the Boltzmann equations in the presence of scalar
perturbations using the line of sight technique \cite{seljak1996line,zaldarriaga1997allsky,lin2006beginners} are given by
\begingroup\small
\begin{gather}
	\frac{d}{d\eta}\Delta_P^{+(S)}(\mathbf{K},k_c)+iK\mu\,\Delta_P^{+(S)}(\mathbf{K},k_c)
	=\dot{\tau}_{e\gamma}\Big[-\Delta_P^{+(S)}(\mathbf{K},k_c)
	+\tfrac{1}{2}\big(1-P_2(\mu)\big)\Pi(\mathbf{K})
	+i\,\frac{\dot{\tau}_{\gamma\gamma'}}{\dot{\tau}_{e\gamma}}
	\big[\Delta_V^{(S)}(\mathbf{K},k_c)*\delta^{\ast}_{2,2}(\mathbf{K})\big]\Big],
	\label{BE+}\\
	\frac{d}{d\eta}\Delta_P^{-(S)}(\mathbf{K},k_c)+iK\mu\,\Delta_P^{-(S)}(\mathbf{K},k_c)
	=\dot{\tau}_{e\gamma}\Big[-\Delta_P^{-(S)}(\mathbf{K},k_c)
	+\tfrac{1}{2}\big(1-P_2(\mu)\big)\Pi(\mathbf{K})
	-i\,\frac{\dot{\tau}_{\gamma\gamma'}}{\dot{\tau}_{e\gamma}}
	\big[\Delta_V^{(S)}(\mathbf{K},k_c)*\delta_{2,2}(\mathbf{K})\big]\Big],
	\label{BE-}\\
	\frac{d}{d\eta}\Delta_V^{(S)}(\mathbf{K},k_c)+iK\mu\,\Delta_V^{(S)}(\mathbf{K},k_c)
	=\dot{\tau}_{e\gamma}\Big[-\Delta_V^{(S)}(\mathbf{K},k_c)
	+\tfrac{3}{2}\mu\,\Delta_{V1}^{(S)}(\mathbf{K},k_c)
	-\frac{\dot{\tau}_{\gamma\gamma'}}{\dot{\tau}_{e\gamma}}\,
	\Delta_Q(\mathbf{K},k_c)*\mathrm{Im}\big(\delta_{2,2}(\mathbf{K})\big)\Big],
	\label{BE3}
\end{gather}
\endgroup
with $ \ast $ denoting the convolution in the Fourier space. $ \dot{\tau}_{e\gamma}(\eta)=a(\eta)n_{e}x_e \sigma_{T} $ is the Thomson scattering optical depth, where $ \tau_{e\gamma}(\eta)=\int_{\eta}^{\eta_{0}}\dot{\tau}_{e\gamma}(\eta')\,d\eta' $ gives the total optical depth. Here $ P_2(\mu) = (3\mu^2-1)/2 $ is the second Legendre polynomial and $ \Pi=\Delta_{I2}+\Delta_{Q0}+\Delta_{Q2} $, with $ a(\eta) $ the scale factor and $ \sigma_{T}=(8\pi/3)\alpha^2/m_{e}^2 $ the Thomson scattering cross-section. The ratio of
the photon--dark-photon to Thomson scattering rates appearing in Eqs.~(\ref{BE+})--(\ref{BE3})
is given explicitly by
\begin{equation}
	\frac{\dot{\tau}_{\gamma \gamma'}}{\dot{\tau}_{e\gamma}}= \frac{6}{\pi^2}\sqrt{\frac{2\pi}{15}}
	\frac{m_e^2 (C_{1}-C_{2}) }{\alpha^2 \Lambda^4}
	\frac{m^2_{\gamma'} n_{\gamma'} k^0_c}{T_{\gamma'}\,a(\eta)\,n_e x_e}.
	\label{eq:ratio}
\end{equation}
By integrating the Boltzmann equation \eqref{BE3} along the line of sight, $ \Delta^{ (S)}_{V}({\bf{K}},\mu,\eta_{0}) $ is given as 
 \begin{eqnarray}\label{delv}
 	\Delta^{ (S)}_{V}({\bf{K}},\mu,\eta_{0})&=&\int_{0}^{\eta_{0}} d\eta\,\dot{\tau}_{e\gamma}\, e^{ix\mu-\tau_{e\gamma}}\,\bigg[\frac{3}{2}\mu\Delta^{(S)}_{V1} -\frac{\dot{\tau}_{\gamma \gamma'}}{\dot{\tau}_{e\gamma}} \Delta_{P}(\mathbf{K},k_{c})\ast \textrm{Im}(\delta_{2,2}(\mathbf{K})) \bigg],
 \end{eqnarray}
 where $ x=K(\eta-\eta_{0}) $ and
 \begin{equation}\label{delp2}
 	\Delta^{(S)}_{P}({\bf{K}},\mu,\eta_{0})=\int_{0}^{\eta_{0}} d\eta\,\dot{\tau}_{e\gamma}\, e^{ix\mu-\tau_{e\gamma}}\,\bigg[\dfrac{3}{4}(1-\mu^{2})\,\Pi(\mathbf{K},\eta)\bigg].
 \end{equation}
The contribution from all the Fourier modes at the direction $ \hat{n} $ is given by,
\begin{eqnarray}\label{allmodes}
	\Delta^{(S)}_{V}(\hat{\bf{n}})=\int d^3{\bf{K}}\,\xi({\bf{K}})\Delta^{(S)}_{V}({\bf{K}},\mu,\eta_{0}),
\end{eqnarray}
where $ \xi({\bf{K}}) $ is the random field characterizing the primordial scalar
perturbation. The spherical-harmonic coefficients of Eq.~(\ref{allmodes}) are
\begin{eqnarray}\label{ClV1}
	a_{V,\ell m}&=&\int\,d\Omega\,Y^{\ast}_{\ell m}(\hat{n})\Delta_{V}(\hat{n}).
\end{eqnarray}
so that the circular polarization power spectrum follows as
\begin{equation}\label{ClV}
	C_{V\ell}^{(S)}=\frac{1}{2\ell+1}\sum_{m}\left\langle a^{\ast}_{V,\ell m} a_{V,\ell m} \right\rangle,
\end{equation}
Substituting Eq.~(\ref{delv}) into Eq.~(\ref{ClV1}) and then into Eq.~(\ref{ClV}), the V-mode power spectrum reads
\begin{align}
	C^{VV}_{\ell} =&\frac{1}{2\ell+1} \int d^3\mathbf{K}\,P_{\phi}({\bf{K}}) \nonumber\\
	& \times \sum_m \left|\int d\Omega_n \, Y^\ast_{\ell m}(\hat{n}) \int_0^{\eta_0}  d\eta\, \dot{\tau}_{e\gamma} \, e^{ix\mu-\tau_{e\gamma}} \left\{\frac{3}{2} \mu  \Delta_{V1}^{(S)} (\mathbf{K}) - \frac{\dot{\tau}_{\gamma \gamma'}}{\dot{\tau}_{e\gamma}}   \textrm{Im} \left[ \delta_{2,2}(\mathbf{K}) \, * \,\Delta_P(\mathbf{K},\mu)  \right]\right\} \right|^2 \, ,  \label{eq:lv}
\end{align}
with $ P_{\phi}({\bf{K}}) $ denoting the primordial power spectrum of scalar perturbations, which is defined as
\begin{eqnarray}\label{ClV3}
	\big\langle\xi(\bf{K})\xi(\bf{K'})\big\rangle=\delta({\bf{K'}-\bf{K}})P_{\phi}({\bf{K}}).
\end{eqnarray}
In Eq.~(\ref{eq:lv}), the first term is negligible compared to the
linear-polarization-sourced term. Keeping only the latter, we obtain an
order-of-magnitude estimate of the circular polarization by introducing the
dimensionless parameter $\beta \equiv \dot{\tau}_{\gamma\gamma'}/\dot{\tau}_{e\gamma}$,
given explicitly by Eq.~(\ref{eq:ratio}), which measures the effective
photon--dark-photon scattering rate relative to the Thomson scattering rate. In
Eq.~(\ref{eq:lv}) this parameter multiplies the visibility function
$\dot{\tau}_{e\gamma}e^{-\tau_{e\gamma}}$, which is the weight with which it enters the
line-of-sight integral. Denoting by $\bar{\beta}_{\rm ave}$ the value of $\beta$ averaged
with this weight, the circular polarization power spectrum
can be estimated relative to the linear polarization power spectrum as follows
\begin{equation}
C^{VV}_{\ell}\leq \bar{\beta}_{ave}^2 \delta_{2,2}^2 C^{EE}_{\ell}
\label{eq:VVbound}
\end{equation}
(the full reduction of Eq.~(\ref{eq:lv}) to spherical Bessel functions, and the reason it
cannot be closed further than Eq.~(\ref{eq:VVbound}), are given in
Appendix~\ref{appA}), where
\begin{equation}
	\bar{\beta}_{\rm ave}\equiv\int_{0}^{\eta_0}d\eta\;\dot{\tau}_{e\gamma}\,e^{-\tau_{e\gamma}}\;\beta(\eta)
	=\int_{0}^{\eta_0}d\eta\;e^{-\tau_{e\gamma}}\;\dot{\tau}_{\gamma\gamma'}(\eta).
	\label{eq:betaave}
\end{equation}
The second form follows from $\beta=\dot{\tau}_{\gamma\gamma'}/\dot{\tau}_{e\gamma}$ and shows
that the Thomson scattering rate cancels, so that $\bar{\beta}_{\rm ave}$ does not depend on the
ionization history after recombination, but only on the fact that $e^{-\tau_{e\gamma}}$
vanishes before recombination and is close to unity afterwards. Inserting
Eq.~(\ref{eq:ratio}) for $\beta$, using $\dot{\tau}_{e\gamma}\,d\eta=a\,n_e x_e\sigma_T\,d\eta$
with $\sigma_T=8\pi\alpha^2/(3m_e^2)$, which cancels $n_e x_e$ in the denominator of
Eq.~(\ref{eq:ratio}), replacing $e^{-\tau_{e\gamma}}$ by a step function at $z_{\rm rec}$,
and changing variables to redshift, $d\eta=dz/H(z)$, we obtain
\begin{equation}
	\bar{\beta}_{\rm ave}=\frac{16}{\pi}\sqrt{\frac{2\pi}{15}}\;
	\frac{(C_{1}-C_{2})\,\bar{k}^{0}_{c}}{\Lambda^4}
	\int_{0}^{z_{\rm rec}}\frac{m^2_{\gamma'}\,n_{\gamma'}(z)}{T_{\gamma'}(z)\,H(z)}\,dz ,
	\label{eq:master}
\end{equation}
where $\bar{k}^{0}_{c}=\pi^4T^{0}_{CMB}/(30\zeta(3))$ is the comoving mean energy of CMB
photons. Everything about the dark photon population enters through the combination
$m^2_{\gamma'}n_{\gamma'}(z)/T_{\gamma'}(z)$ under the integral, i.e.\ through its number
density, its temperature or velocity dispersion, and their evolution with redshift,
while all of the particle-physics input is contained in the prefactor
$(C_{1}-C_{2})/\Lambda^4$.
\section{Dark Photon Production Scenarios and the Parameter \texorpdfstring{$\beta$}{beta}}\label{V}

The magnitude of the circular polarization is controlled by $\bar{\beta}_{\rm ave}$,
Eq.~(\ref{eq:master}), through the redshift dependence of
$m^2_{\gamma'}n_{\gamma'}/T_{\gamma'}$, which is fixed once the production mechanism of
the dark photons is specified \cite{fabbrichesi2020dark}. We consider a
thermal relic, whose momentum distribution is inherited from the Standard Model plasma,
and a cold population produced by the decay of a heavy parent, whose temperature is set
by its velocity.
As the sensitivity of forthcoming experiments is not yet established, we adopt
$\ell(\ell+1)C_\ell^{VV}/(2\pi)\gtrsim0.01\,\mathrm{nK}^2$ as an order-of-magnitude
benchmark for observability, with $|\delta_{2,2}|\sim10^{-5}$ and
$C_\ell^{EE}\sim0.1\,(\mu\mathrm{K})^2$ in Eq.~(\ref{eq:VVbound}).

\subsection{Thermal Relic}\label{sec:thermal}

If dark photons were in thermal equilibrium with the Standard Model bath, through the
interaction of Eq.~(\ref{eq:lint}) itself or through additional dark-sector
interactions, and decoupled while relativistic, their momentum distribution is frozen
from then on, with temperature
\begin{equation}
	T_{\gamma'}(z)=r\,T^{0}_{CMB}(1+z),\qquad
	r=\left[\frac{g_{*s}(T_0)}{g_{*s}(T_{\rm dec})}\right]^{1/3},
	\label{eq:Tthermal}
\end{equation}
where $r$ ranges from $0.33$ for decoupling above the electroweak scale to $1$ for
decoupling after $e^{\pm}$ annihilation, and their number density is conserved,
\begin{equation}
	n_{\gamma'}(z) = \frac{\zeta(3)}{\pi^2}\,g_{\gamma'}\,T_{\gamma'}^3(z)
	\simeq 616\,r^3(1+z)^3~\mathrm{cm}^{-3},
	\label{eq:n_thermal_rel}
\end{equation}
with $g_{\gamma'}=3$ for a massive vector boson. As shown in Sec.~\ref{III}, the source of
circular polarization vanishes for relativistic dark photons, so a thermal relic
contributes only if it is non-relativistic at recombination,
$m_{\gamma'}\gtrsim\langle p^2\rangle^{1/2}\simeq3\,T_{\gamma'}(z_{\rm rec})$. Its kinetic temperature, the
quantity entering the Maxwell--Boltzmann distribution of Eq.~(\ref{eq:rho1}), is then
\begin{equation}
	T_{\gamma'}(z)=\frac{\langle p^2\rangle}{3m_{\gamma'}}
	=\frac{4\zeta(5)}{\zeta(3)}\,\frac{r^2\,(T^{0}_{CMB})^2}{m_{\gamma'}}\,(1+z)^2
	\simeq3.45\,\frac{r^2\,(T^{0}_{CMB})^2}{m_{\gamma'}}\,(1+z)^2 ,
	\label{eq:Tkin}
\end{equation}
so that
\begin{equation}
	\frac{m^2_{\gamma'}n_{\gamma'}(z)}{T_{\gamma'}(z)}
	=\frac{m^3_{\gamma'}\,n^{0}_{\gamma'}}{3.45\,r^2\,(T^{0}_{CMB})^2}\,(1+z).
	\label{eq:combo_nr}
\end{equation}
With the matter-era expansion rate $H(z)\simeq H_0\sqrt{\Omega_m}\,(1+z)^{3/2}$, valid over
the whole integration range, the integral in Eq.~(\ref{eq:master}) gives
\begin{equation}
	\bar{\beta}_{\rm ave}\simeq 4\times10^{27}\,r\left(\frac{m_{\gamma'}}{\mathrm{eV}}\right)^{3}
	\frac{(C_{1}-C_{2})}{\Lambda^4},
	\label{thermalmassive}
\end{equation}
where we use $T^{0}_{CMB}=2.35\times10^{-4}$~eV, $z_{\rm rec}=1100$ and the Planck values of
$H_0$ and $\Omega_m$ \cite{planck2018params}, and express $\Lambda$ in eV here and below.
Two observational bounds restrict $r$ and $m_{\gamma'}$. The Planck limit
$\Delta N_{\rm eff}\lesssim0.3$ \cite{planck2018params} requires $r\lesssim0.4$, i.e.\
decoupling before the QCD transition, and we take $r=0.33$, the value for decoupling
above the electroweak scale. The relic also contributes a hot component to the dark
matter, and the Planck bound on such a component, the same as for massive neutrinos
\cite{planck2018params},
\begin{equation}
	\Omega_{\gamma'}h^2=0.058\,r^3\left(\frac{m_{\gamma'}}{\mathrm{eV}}\right)\lesssim1.3\times10^{-3},
	\label{eq:hotDM}
\end{equation}
requires $m_{\gamma'}r^3\lesssim0.02$~eV, i.e.\ $m_{\gamma'}\lesssim0.6$~eV, so that the
relic makes up at most $1\%$ of the dark matter. For $r=0.33$ and $m_{\gamma'}=0.6$~eV,
Eq.~(\ref{thermalmassive}) gives
$\bar{\beta}_{\rm ave}\simeq3\times10^{26}(C_{1}-C_{2})/\Lambda^4$, and Eq.~(\ref{eq:VVbound})
shows that the benchmark is reached for $\Lambda/(C_{1}-C_{2})^{1/4}\lesssim1.8$~MeV.
\subsection{Decay-Produced Cold Population}\label{sec:nonthermal}

A larger $\bar{\beta}_{\rm ave}$ requires a population that is both denser and colder at
recombination than a thermal relic. This is the case for dark photons produced
non-thermally through the decay $X\to\gamma'\gamma'$ of a non-relativistic parent of mass
$M$ at redshift $z_d$. Each dark photon is born with momentum $M/2$ (for
$M\gg2m_{\gamma'}$), which subsequently redshifts, so that once it is non-relativistic
its velocity is
\begin{equation}
	v(z)=v_0\,(1+z),\qquad v_0\equiv v(z=0)=\frac{M}{2m_{\gamma'}}\,\frac{1}{1+z_d}.
	\label{eq:vz}
\end{equation}
The population is characterized by $v_0$ and by the fraction $f_{\rm DM}$ of the dark
matter density it carries, $\rho_{\gamma'}(z)=f_{\rm DM}\,\rho_{\rm DM}(z)=m_{\gamma'}n_{\gamma'}(z)$,
and its kinetic temperature is $T_{\gamma'}=\langle p^2\rangle/3m_{\gamma'}=\frac{1}{3}m_{\gamma'}v^2$.
The dependence on the dark photon mass cancels,
\begin{equation}
	\frac{m^2_{\gamma'}n_{\gamma'}(z)}{T_{\gamma'}(z)}
	=\frac{3f_{\rm DM}\,\rho_{\rm DM}(z)}{v(z)^2}
	=\frac{3f_{\rm DM}\,\rho^{0}_{\rm DM}}{v_0^2}\,(1+z),
	\label{eq:combo_nt}
\end{equation}
with the same redshift dependence as Eq.~(\ref{eq:combo_nr}). The non-relativistic
treatment requires $v(z_{\rm rec})=(1+z_{\rm rec})\,v_0\ll1$, i.e.\ $v_0\lesssim10^{-4}$,
which by Eq.~(\ref{eq:vz}) requires the decay to occur deep in the radiation era. A
component with this velocity behaves as warm dark matter, and bounds on mixed
cold-plus-warm dark matter from structure formation restrict it to a subdominant
fraction, which motivates the choice $f_{\rm DM}=0.1$. With the same integral as above
and $\rho^{0}_{\rm DM}=\Omega_{\rm DM}\rho_c=1.39\times10^3$~eV/cm$^3$,
\begin{equation}
	\bar{\beta}_{\rm ave}\simeq 5.4\times10^{28}\left(\frac{f_{\rm DM}}{0.1}\right)
	\left(\frac{10^{-4}}{v_0}\right)^{2}\frac{(C_{1}-C_{2})}{\Lambda^4},
	\label{nonthermallambda}
\end{equation}
and the benchmark is reached for $\Lambda/(C_{1}-C_{2})^{1/4}\lesssim6.5$~MeV. Since
$\bar{\beta}_{\rm ave}\propto f_{\rm DM}/v_0^2$, the required scale varies only as
$(f_{\rm DM}/v_0^2)^{1/4}$. A colder or more abundant population raises the scale
accordingly; for $f_{\rm DM}=0.3$ and $v_0=10^{-5}$, still compatible with structure
formation, the benchmark is reached for $\Lambda/(C_{1}-C_{2})^{1/4}\lesssim27$~MeV, for the
same quadrupole $\delta_{2,2}$.
The growth as $v_0^{-2}$ cannot, however, be extrapolated to
arbitrarily small $v_0$, because in that limit the population approaches a coherent condensate,
such as dark photon dark matter from the misalignment mechanism or from inflationary
fluctuations \cite{fabbrichesi2020dark,caputo2021dark}, for which the description of the
anisotropy $\delta(\mathbf{x},\hat{q}')$ as a perturbation of the Maxwell--Boltzmann
distribution in Eq.~(\ref{eq:rho1}) no longer applies.

The two populations thus share the redshift dependence
$m^2_{\gamma'}n_{\gamma'}/T_{\gamma'}\propto(1+z)$ and differ only in their energy density
and velocity at recombination. The thermal relic is limited to $1\%$ of the dark matter, whereas the decay-produced population can be more
abundant and colder, $v(z_{\rm rec})\simeq0.1$, which is why it gives the larger
$\bar{\beta}_{\rm ave}$. The required effective scale is about $1.8$~MeV in the first case
and about $6.5$~MeV in the second.

\section{Conclusion}\label{VI}

In this work we have investigated the generation of circular polarization in the CMB
through forward scattering between ordinary photons and a cosmological dark-photon
background. The minimal kinetic-mixing portal induces direct couplings of the dark photon
to Standard Model fermions that are so strongly constrained by stellar, laboratory and
cosmological bounds that the resulting $V$-mode is unobservably small. We therefore
adopted a model-independent effective field theory in which the two sectors communicate
only through dimension-8 Euler--Heisenberg-type operators, which couple to field
strengths and not to fermion currents, with a scale $\Lambda$ and Wilson coefficients
$C_{1,2}$ as free parameters. Using the quantum Boltzmann equation for the photon density
matrix, we derived the evolution equations for the Stokes parameters and found that a
polarized dark-photon background, with a quadrupolar asymmetry between its transverse and
longitudinal polarizations, converts the linear polarization of the CMB into circular
polarization. The source is proportional to $m^2_{\gamma'}n_{\gamma'}/T_{\gamma'}$, where
the factor $m^2_{\gamma'}$ reflects the longitudinal polarization of the massive dark
photon, and it vanishes in the relativistic limit, in which the longitudinal mode
decouples.

Incorporating the scattering term into the line-of-sight solution for scalar
perturbations, we estimated the circular polarization power spectrum as
$C_\ell^{VV}\lesssim\bar{\beta}_{\rm ave}^{\,2}\,|\delta_{2,2}|^2\,C_\ell^{EE}$, where
$\delta_{2,2}$ is the dark-photon quadrupole and $\bar{\beta}_{\rm ave}$ is the
photon--dark-photon scattering rate integrated over conformal time since
recombination. We evaluated $\bar{\beta}_{\rm ave}$ for the two classes of populations, taking
$\ell(\ell+1)C_\ell^{VV}/2\pi\gtrsim0.01\,\mathrm{nK}^2$ as an order-of-magnitude benchmark
with $|\delta_{2,2}|\sim10^{-5}$. A thermal
relic that decoupled while relativistic is constrained by $N_{\rm eff}$ and by the hot-dark-matter bound to sub-eV masses, in which it makes up at most $1\%$ of the dark matter; there the
benchmark requires $\Lambda/(C_1-C_2)^{1/4}\lesssim1.8$~MeV. A cold population produced
by the decay of a heavy parent is characterized by its dark-matter fraction and its
velocity, with the mass dependence cancelling in
$m_{\gamma'}^2n_{\gamma'}/T_{\gamma'}=3f_{\rm DM}\rho_{\rm DM}/v^2$; for $f_{\rm DM}=0.1$ and a
present velocity $v_0=10^{-4}$, i.e.\ $v\simeq0.1$ at recombination, the benchmark is
reached for $\Lambda/(C_1-C_2)^{1/4}\lesssim6.5$~MeV, rising to about $27$~MeV
for a colder and more abundant population. The two populations share the
redshift dependence $m_{\gamma'}^2n_{\gamma'}/T_{\gamma'}\propto(1+z)$ and differ only in
their energy density and velocity at recombination, which is why the decay-produced
population, more abundant and colder, is the more favourable route to an observable
signal in this framework. Existing bounds on kinetic mixing do not apply to a pure field-strength interaction,
and an order-of-magnitude estimate of stellar energy loss through
$\gamma\gamma\to\gamma'\gamma'$ suggests $\Lambda\gtrsim\mathcal{O}(10\,\mathrm{MeV})$ for
Wilson coefficients of order unity. A dedicated calculation of the
astrophysical constraints on dimension-8 photon--dark-photon operators is an important
direction for future work.

Our results establish CMB circular polarization as a probe of this class of interactions
that is complementary to laboratory and astrophysical searches. Whereas those constrain
$\Lambda$ through high-energy scattering or energy-loss rates, the $V$-mode depends on
$\Lambda$ together with the phase-space density of the dark-photon background, and is
therefore sensitive to its mass, its abundance and its thermal state at recombination in
a way that no terrestrial experiment is. A detection of primordial circular polarization
by future high-sensitivity polarimeters would point to a massive dark photon with a
cold and abundant cosmological population and an interaction scale in the MeV range.

\section{ACKNOWLEDGEMENTS}

The scattering amplitude in Eq.~(\ref{eq:amplitude}) was calculated using the GRTensorII
package.

\appendix
\section{Power Spectrum in Terms of Spherical Bessel Functions}\label{appA}

In this appendix we carry the line-of-sight result of Eq.~(\ref{eq:lv}) further than the
order-of-magnitude estimate used in the main text, reducing it to spherical Bessel
functions using the standard line-of-sight technique
\cite{seljak1996line,zaldarriaga1997allsky}. As in Sec.~\ref{IV}, we drop the free-streaming term of
Eq.~(\ref{eq:lv}), which is negligible compared to the linear-polarization-sourced term,
and substitute Eq.~(\ref{delp2}) for $\Delta_P$. Since the product of the two visibility
functions this generates is sharply peaked at $\eta'\simeq\eta$, Eq.~(\ref{eq:lv}) reduces
to
\begin{equation}
	C_\ell^{VV} \simeq \frac{1}{2\ell+1}\int d^3\mathbf{K}\,P_{\phi}(\mathbf{K})
	\sum_m\left|\int d\Omega_n\,Y^*_{\ell m}(\hat{n})\int_0^{\eta_0}d\eta\,
	\dot{\tau}_{e\gamma}e^{-\tau_{e\gamma}}\,\frac{3}{4}\beta(1-\mu^2)e^{ix\mu}\,
	\mathrm{Im}\big[\delta_{2,2}(\mathbf{K})*\Pi(\mathbf{K},\eta)\big]\right|^2.
	\label{eq:appA2}
\end{equation}
Using the projection integral \cite{schmidt2012cosmic}
\begin{equation}
	\int d\Omega\,Y^*_{\ell m}(\theta,\phi)\,e^{ix\mu}e^{ir\phi}(1-\mu^2)^{|r|/2}
	=\sqrt{4\pi(2\ell+1)}\sqrt{\frac{(\ell+|r|)!}{(\ell-|r|)!}}\,i^r i^\ell\,
	\frac{j_\ell(x)}{x^{|r|}}\,\delta_{mr},
	\label{eq:appProj}
\end{equation}
together with the spherical Bessel differential equation and recurrence relation, the
angular integral in Eq.~(\ref{eq:appA2}) can be performed, giving
\begin{equation}
	C_\ell^{VV}=(4\pi)(\ell^4-2\ell^3+\ell^2)\int d^3\mathbf{K}\,P_{\phi}(\mathbf{K})
	\left|\int_0^{\eta_0}d\eta\,\dot{\tau}_{e\gamma}e^{-\tau_{e\gamma}}\,\frac{3}{4}\beta
	\left[\frac{2j_{\ell+1}(x)}{\ell(\ell-1)x}+\frac{j_\ell(x)}{x^2}\right]
	\mathrm{Im}\big[\delta_{2,2}(\mathbf{K})*\Pi(\mathbf{K})\big]\right|^2.
	\label{eq:appBessel}
\end{equation}
Since the dark-photon quadrupole is sourced by the same scalar perturbations treated
throughout Sec.~\ref{IV}, it is axially symmetric about $\mathbf{K}-\mathbf{K}_1$, so that
$\delta_{2,2}(\mathbf{K}-\mathbf{K}_1)$ is related to $\delta_{2,0}(|\mathbf{K}-\mathbf{K}_1|)$,
evaluated in the frame aligned with $\mathbf{K}-\mathbf{K}_1$, through the Wigner rotation
matrix \cite{wigner2012group}
\begin{align}
	\delta_{2,2}(\mathbf{K}-\mathbf{K}_1)&=D^2_{2,0}(\phi_{K-K_1},\theta_{K-K_1},0)\,
	\delta_{2,0}(|\mathbf{K}-\mathbf{K}_1|) \notag\\
	&=\frac{\sqrt{6}}{4}\sin^2\theta_{K-K_1}\,e^{-2i\phi_{K-K_1}}\,
	\delta_{2,0}(|\mathbf{K}-\mathbf{K}_1|).
	\label{eq:appRot}
\end{align}
This reduces the convolution in Eq.~(\ref{eq:appBessel}) to
\begin{equation}
	\mathrm{Im}\big[\delta_{2,2}(\mathbf{K})*\Pi(\mathbf{K},\eta)\big]
	=-\frac{\sqrt{6}}{4}\int\frac{d^3\mathbf{K}_1}{(2\pi)^3}\,
	\sin^2\theta_{K-K_1}\sin(2\phi_{K-K_1})\,\delta_{2,0}(|\mathbf{K}-\mathbf{K}_1|)\,
	\Pi(\mathbf{K}_1),
	\label{eq:appConv}
\end{equation}
so that the $V$-mode power spectrum takes the closed form
\begin{align}
	C_\ell^{VV}=(4\pi)(\ell^4-2\ell^3+\ell^2)\int d^3\mathbf{K}\,P_{\phi}(\mathbf{K})
	\bigg|\int_0^{\eta_0}d\eta\,\dot{\tau}_{e\gamma}e^{-\tau_{e\gamma}}\,
	\frac{3\sqrt{6}}{16}\beta
	\left[\frac{2j_{\ell+1}(x)}{\ell(\ell-1)x}+\frac{j_\ell(x)}{x^2}\right]
	\notag\\
	\times\int\frac{d^3\mathbf{K}_1}{(2\pi)^3}\,
	\sin^2\theta_{K-K_1}\sin(2\phi_{K-K_1})\,\delta_{2,0}(|\mathbf{K}-\mathbf{K}_1|)\,
	\Pi(\mathbf{K}_1)\bigg|^2.
	\label{eq:appFinal}
\end{align}
To compare Eq.~(\ref{eq:appFinal}) against the observable linear-polarization power
spectrum, we also record $C_\ell^{EE}$, obtained by the same line-of-sight calculation,
which reduces to the standard result
\begin{equation}
	C_\ell^{EE}=(4\pi)(\ell^4+2\ell^3-\ell^2-2\ell)\int d^3\mathbf{K}\,P_{\phi}(\mathbf{K})
	\left|\int_0^{\eta_0}d\eta\,\dot{\tau}_{e\gamma}e^{-\tau_{e\gamma}}\,\frac{3}{4}
	\Pi(\mathbf{K},\eta)\,\frac{j_\ell(x)}{x^2}\right|^2.
	\label{eq:appEE}
\end{equation}
Closing Eq.~(\ref{eq:appFinal}) into an analytic bound purely in terms of $C_\ell^{EE}$
would require the power spectrum $\langle\delta_{2,0}\delta^*_{2,0}\rangle$ of the
dark-photon quadrupole itself. Unlike the CMB's own quadrupole, this is not fixed by any
existing measurement. It depends on the phase-space distribution and free-streaming
history of the dark-photon population, which are exactly the production-mechanism-dependent
quantities discussed in Sec.~\ref{V}. Lacking an independent determination of this power
spectrum, we instead estimate $|\delta_{2,2}|$ by order of magnitude and use it to bound
$C_\ell^{VV}$ through Eq.~(\ref{eq:VVbound}). In the absence of a detailed Boltzmann
hierarchy for the dark photon sector, we make this estimate explicit by parametrizing
its magnitude as
\begin{equation}
	|\delta_{2,2}| \equiv \chi\,\Phi,
	\label{eq:quad_param}
\end{equation}
where $\Phi \sim 10^{-5}$ is the primordial gravitational potential and $\chi \leq 1$
encodes possible suppression due to late production or reduced free streaming. For a
population that is present well before recombination and shares the primordial
perturbations, one expects $\chi = \mathcal{O}(1)$.

\bibliography{references.bib}

\begin{thebibliography}{42}%
\makeatletter
\providecommand \@ifxundefined [1]{%
 \@ifx{#1\undefined}
}%
\providecommand \@ifnum [1]{%
 \ifnum #1\expandafter \@firstoftwo
 \else \expandafter \@secondoftwo
 \fi
}%
\providecommand \@ifx [1]{%
 \ifx #1\expandafter \@firstoftwo
 \else \expandafter \@secondoftwo
 \fi
}%
\providecommand \natexlab [1]{#1}%
\providecommand \enquote  [1]{``#1''}%
\providecommand \bibnamefont  [1]{#1}%
\providecommand \bibfnamefont [1]{#1}%
\providecommand \citenamefont [1]{#1}%
\providecommand \href@noop [0]{\@secondoftwo}%
\providecommand \href [0]{\begingroup \@sanitize@url \@href}%
\providecommand \@href[1]{\@@startlink{#1}\@@href}%
\providecommand \@@href[1]{\endgroup#1\@@endlink}%
\providecommand \@sanitize@url [0]{\catcode `\\12\catcode `\$12\catcode
  `\&12\catcode `\#12\catcode `\^12\catcode `\_12\catcode `\%12\relax}%
\providecommand \@@startlink[1]{}%
\providecommand \@@endlink[0]{}%
\providecommand \url  [0]{\begingroup\@sanitize@url \@url }%
\providecommand \@url [1]{\endgroup\@href {#1}{\urlprefix }}%
\providecommand \urlprefix  [0]{URL }%
\providecommand \Eprint [0]{\href }%
\providecommand \doibase [0]{http://dx.doi.org/}%
\providecommand \selectlanguage [0]{\@gobble}%
\providecommand \bibinfo  [0]{\@secondoftwo}%
\providecommand \bibfield  [0]{\@secondoftwo}%
\providecommand \translation [1]{[#1]}%
\providecommand \BibitemOpen [0]{}%
\providecommand \bibitemStop [0]{}%
\providecommand \bibitemNoStop [0]{.\EOS\space}%
\providecommand \EOS [0]{\spacefactor3000\relax}%
\providecommand \BibitemShut  [1]{\csname bibitem#1\endcsname}%
\let\auto@bib@innerbib\@empty
\bibitem [{\citenamefont {Hu}\ and\ \citenamefont
  {Dodelson}(2002)}]{hu2002cmb}%
  \BibitemOpen
  \bibfield  {author} {\bibinfo {author} {\bibfnamefont {W.}~\bibnamefont
  {Hu}}\ and\ \bibinfo {author} {\bibfnamefont {S.}~\bibnamefont {Dodelson}},\
  }\href {\doibase 10.1146/annurev.astro.40.060401.093926} {\bibfield
  {journal} {\bibinfo  {journal} {Annual Review of Astronomy and Astrophysics}\
  }\textbf {\bibinfo {volume} {40}},\ \bibinfo {pages} {171} (\bibinfo {year}
  {2002})},\ \Eprint {http://arxiv.org/abs/astro-ph/0110414}
  {arXiv:astro-ph/0110414} \BibitemShut {NoStop}%
\bibitem [{\citenamefont {{Kosowsky}}(1996)}]{1996AnPhy.246...49K}%
  \BibitemOpen
  \bibfield  {author} {\bibinfo {author} {\bibfnamefont {A.}~\bibnamefont
  {{Kosowsky}}},\ }\href {\doibase 10.1006/aphy.1996.0020} {\bibfield
  {journal} {\bibinfo  {journal} {Annals of Physics}\ }\textbf {\bibinfo
  {volume} {246}},\ \bibinfo {pages} {49} (\bibinfo {year} {1996})},\ \Eprint
  {http://arxiv.org/abs/astro-ph/9501045} {arXiv:astro-ph/9501045 [astro-ph]}
  \BibitemShut {NoStop}%
\bibitem [{\citenamefont {Chandrasekhar}(1960)}]{chandrasekhar_radiative_1960}%
  \BibitemOpen
  \bibfield  {author} {\bibinfo {author} {\bibfnamefont {S.}~\bibnamefont
  {Chandrasekhar}},\ }\href@noop {} {\emph {\bibinfo {title} {Radiative
  Transfer}}}\ (\bibinfo  {publisher} {Dover Publications},\ \bibinfo {address}
  {New York},\ \bibinfo {year} {1960})\BibitemShut {NoStop}%
\bibitem [{\citenamefont {Seljak}\ and\ \citenamefont
  {Zaldarriaga}(1996)}]{seljak1996line}%
  \BibitemOpen
  \bibfield  {author} {\bibinfo {author} {\bibfnamefont {U.}~\bibnamefont
  {Seljak}}\ and\ \bibinfo {author} {\bibfnamefont {M.}~\bibnamefont
  {Zaldarriaga}},\ }\href {\doibase 10.1086/177793} {\bibfield  {journal}
  {\bibinfo  {journal} {The Astrophysical Journal}\ }\textbf {\bibinfo {volume}
  {469}},\ \bibinfo {pages} {437} (\bibinfo {year} {1996})},\ \Eprint
  {http://arxiv.org/abs/astro-ph/9603033} {arXiv:astro-ph/9603033} \BibitemShut
  {NoStop}%
\bibitem [{\citenamefont {Zaldarriaga}\ and\ \citenamefont
  {Seljak}(1997)}]{zaldarriaga1997allsky}%
  \BibitemOpen
  \bibfield  {author} {\bibinfo {author} {\bibfnamefont {M.}~\bibnamefont
  {Zaldarriaga}}\ and\ \bibinfo {author} {\bibfnamefont {U.}~\bibnamefont
  {Seljak}},\ }\href {\doibase 10.1103/PhysRevD.55.1830} {\bibfield  {journal}
  {\bibinfo  {journal} {Physical Review D}\ }\textbf {\bibinfo {volume} {55}},\
  \bibinfo {pages} {1830} (\bibinfo {year} {1997})},\ \Eprint
  {http://arxiv.org/abs/astro-ph/9609170} {arXiv:astro-ph/9609170} \BibitemShut
  {NoStop}%
\bibitem [{\citenamefont {Kamionkowski}\ \emph {et~al.}(1997)\citenamefont
  {Kamionkowski}, \citenamefont {Kosowsky},\ and\ \citenamefont
  {Stebbins}}]{kamionkowski1997statistics}%
  \BibitemOpen
  \bibfield  {author} {\bibinfo {author} {\bibfnamefont {M.}~\bibnamefont
  {Kamionkowski}}, \bibinfo {author} {\bibfnamefont {A.}~\bibnamefont
  {Kosowsky}}, \ and\ \bibinfo {author} {\bibfnamefont {A.}~\bibnamefont
  {Stebbins}},\ }\href {\doibase 10.1103/PhysRevD.55.7368} {\bibfield
  {journal} {\bibinfo  {journal} {Physical Review D}\ }\textbf {\bibinfo
  {volume} {55}},\ \bibinfo {pages} {7368} (\bibinfo {year} {1997})},\ \Eprint
  {http://arxiv.org/abs/astro-ph/9611125} {arXiv:astro-ph/9611125} \BibitemShut
  {NoStop}%
\bibitem [{\citenamefont {Hu}\ and\ \citenamefont
  {White}(1997)}]{hu1997primer}%
  \BibitemOpen
  \bibfield  {author} {\bibinfo {author} {\bibfnamefont {W.}~\bibnamefont
  {Hu}}\ and\ \bibinfo {author} {\bibfnamefont {M.}~\bibnamefont {White}},\
  }\href {\doibase 10.1016/S1384-1076(97)00022-5} {\bibfield  {journal}
  {\bibinfo  {journal} {New Astronomy}\ }\textbf {\bibinfo {volume} {2}},\
  \bibinfo {pages} {323} (\bibinfo {year} {1997})},\ \Eprint
  {http://arxiv.org/abs/astro-ph/9706147} {arXiv:astro-ph/9706147} \BibitemShut
  {NoStop}%
\bibitem [{\citenamefont {Bennett}\ \emph {et~al.}(2013)\citenamefont
  {Bennett}, \citenamefont {Larson}, \citenamefont {Weiland}, \citenamefont
  {Jarosik}, \citenamefont {Hinshaw}, \citenamefont {Odegard}, \citenamefont
  {Smith}, \citenamefont {Hill}, \citenamefont {Gold}, \citenamefont {Halpern}
  \emph {et~al.}}]{bennett2013wmap9}%
  \BibitemOpen
  \bibfield  {author} {\bibinfo {author} {\bibfnamefont {C.~L.}\ \bibnamefont
  {Bennett}}, \bibinfo {author} {\bibfnamefont {D.}~\bibnamefont {Larson}},
  \bibinfo {author} {\bibfnamefont {J.~L.}\ \bibnamefont {Weiland}}, \bibinfo
  {author} {\bibfnamefont {N.}~\bibnamefont {Jarosik}}, \bibinfo {author}
  {\bibfnamefont {G.}~\bibnamefont {Hinshaw}}, \bibinfo {author} {\bibfnamefont
  {N.}~\bibnamefont {Odegard}}, \bibinfo {author} {\bibfnamefont {K.~M.}\
  \bibnamefont {Smith}}, \bibinfo {author} {\bibfnamefont {R.~S.}\ \bibnamefont
  {Hill}}, \bibinfo {author} {\bibfnamefont {B.}~\bibnamefont {Gold}}, \bibinfo
  {author} {\bibfnamefont {M.}~\bibnamefont {Halpern}},  \emph {et~al.},\
  }\href {\doibase 10.1088/0067-0049/208/2/20} {\bibfield  {journal} {\bibinfo
  {journal} {The Astrophysical Journal Supplement Series}\ }\textbf {\bibinfo
  {volume} {208}},\ \bibinfo {pages} {20} (\bibinfo {year} {2013})},\ \Eprint
  {http://arxiv.org/abs/1212.5225} {arXiv:1212.5225} \BibitemShut {NoStop}%
\bibitem [{\citenamefont {{Planck Collaboration}}\ \emph
  {et~al.}(2020{\natexlab{a}})\citenamefont {{Planck Collaboration}},
  \citenamefont {Aghanim} \emph {et~al.}}]{planck2018params}%
  \BibitemOpen
  \bibfield  {author} {\bibinfo {author} {\bibnamefont {{Planck
  Collaboration}}}, \bibinfo {author} {\bibfnamefont {N.}~\bibnamefont
  {Aghanim}},  \emph {et~al.},\ }\href {\doibase 10.1051/0004-6361/201833910}
  {\bibfield  {journal} {\bibinfo  {journal} {Astronomy \& Astrophysics}\
  }\textbf {\bibinfo {volume} {641}},\ \bibinfo {pages} {A6} (\bibinfo {year}
  {2020}{\natexlab{a}})},\ \Eprint {http://arxiv.org/abs/1807.06209}
  {arXiv:1807.06209} \BibitemShut {NoStop}%
\bibitem [{\citenamefont {{Planck Collaboration}}\ \emph
  {et~al.}(2020{\natexlab{b}})\citenamefont {{Planck Collaboration}},
  \citenamefont {Aghanim} \emph {et~al.}}]{planck2018spectra}%
  \BibitemOpen
  \bibfield  {author} {\bibinfo {author} {\bibnamefont {{Planck
  Collaboration}}}, \bibinfo {author} {\bibfnamefont {N.}~\bibnamefont
  {Aghanim}},  \emph {et~al.},\ }\href {\doibase 10.1051/0004-6361/201936386}
  {\bibfield  {journal} {\bibinfo  {journal} {Astronomy \& Astrophysics}\
  }\textbf {\bibinfo {volume} {641}},\ \bibinfo {pages} {A5} (\bibinfo {year}
  {2020}{\natexlab{b}})},\ \Eprint {http://arxiv.org/abs/1907.12875}
  {arXiv:1907.12875} \BibitemShut {NoStop}%
\bibitem [{\citenamefont {{BICEP/Keck Collaboration}}\ \emph
  {et~al.}(2021)\citenamefont {{BICEP/Keck Collaboration}}, \citenamefont {Ade}
  \emph {et~al.}}]{bicepkeck2021}%
  \BibitemOpen
  \bibfield  {author} {\bibinfo {author} {\bibnamefont {{BICEP/Keck
  Collaboration}}}, \bibinfo {author} {\bibfnamefont {P.~A.~R.}\ \bibnamefont
  {Ade}},  \emph {et~al.},\ }\href {\doibase 10.1103/PhysRevLett.127.151301}
  {\bibfield  {journal} {\bibinfo  {journal} {Physical Review Letters}\
  }\textbf {\bibinfo {volume} {127}},\ \bibinfo {pages} {151301} (\bibinfo
  {year} {2021})},\ \Eprint {http://arxiv.org/abs/2110.00483}
  {arXiv:2110.00483} \BibitemShut {NoStop}%
\bibitem [{\citenamefont {Choi}\ \emph {et~al.}(2020)\citenamefont {Choi},
  \citenamefont {Hasselfield}, \citenamefont {Ho} \emph
  {et~al.}}]{act2020dr4spectra}%
  \BibitemOpen
  \bibfield  {author} {\bibinfo {author} {\bibfnamefont {S.~K.}\ \bibnamefont
  {Choi}}, \bibinfo {author} {\bibfnamefont {M.}~\bibnamefont {Hasselfield}},
  \bibinfo {author} {\bibfnamefont {S.-P.~P.}\ \bibnamefont {Ho}},  \emph
  {et~al.},\ }\href {\doibase 10.1088/1475-7516/2020/12/045} {\bibfield
  {journal} {\bibinfo  {journal} {Journal of Cosmology and Astroparticle
  Physics}\ }\textbf {\bibinfo {volume} {2020}},\ \bibinfo {pages} {045}
  (\bibinfo {year} {2020})},\ \Eprint {http://arxiv.org/abs/2007.07289}
  {arXiv:2007.07289} \BibitemShut {NoStop}%
\bibitem [{\citenamefont {Qu}\ \emph {et~al.}(2024)\citenamefont {Qu},
  \citenamefont {Sherwin}, \citenamefont {Madhavacheril} \emph
  {et~al.}}]{act2023dr6lensing}%
  \BibitemOpen
  \bibfield  {author} {\bibinfo {author} {\bibfnamefont {F.~J.}\ \bibnamefont
  {Qu}}, \bibinfo {author} {\bibfnamefont {B.~D.}\ \bibnamefont {Sherwin}},
  \bibinfo {author} {\bibfnamefont {M.~S.}\ \bibnamefont {Madhavacheril}},
  \emph {et~al.},\ }\href {\doibase 10.3847/1538-4357/acfe06} {\bibfield
  {journal} {\bibinfo  {journal} {The Astrophysical Journal}\ }\textbf
  {\bibinfo {volume} {962}},\ \bibinfo {pages} {112} (\bibinfo {year}
  {2024})},\ \Eprint {http://arxiv.org/abs/2304.05202} {arXiv:2304.05202}
  \BibitemShut {NoStop}%
\bibitem [{\citenamefont {Mainini}\ \emph {et~al.}(2013)\citenamefont
  {Mainini}, \citenamefont {Minelli}, \citenamefont {Gervasi}, \citenamefont
  {Boella}, \citenamefont {Sironi}, \citenamefont {Bau'}, \citenamefont
  {Banfi}, \citenamefont {Passerini}, \citenamefont {De~Lucia},\ and\
  \citenamefont {Cavaliere}}]{mainini2013improved}%
  \BibitemOpen
  \bibfield  {author} {\bibinfo {author} {\bibfnamefont {R.}~\bibnamefont
  {Mainini}}, \bibinfo {author} {\bibfnamefont {D.}~\bibnamefont {Minelli}},
  \bibinfo {author} {\bibfnamefont {M.}~\bibnamefont {Gervasi}}, \bibinfo
  {author} {\bibfnamefont {G.}~\bibnamefont {Boella}}, \bibinfo {author}
  {\bibfnamefont {G.}~\bibnamefont {Sironi}}, \bibinfo {author} {\bibfnamefont
  {A.}~\bibnamefont {Bau'}}, \bibinfo {author} {\bibfnamefont {S.}~\bibnamefont
  {Banfi}}, \bibinfo {author} {\bibfnamefont {A.}~\bibnamefont {Passerini}},
  \bibinfo {author} {\bibfnamefont {A.}~\bibnamefont {De~Lucia}}, \ and\
  \bibinfo {author} {\bibfnamefont {F.}~\bibnamefont {Cavaliere}},\ }\href
  {\doibase 10.1088/1475-7516/2013/08/033} {\bibfield  {journal} {\bibinfo
  {journal} {Journal of Cosmology and Astroparticle Physics}\ }\textbf
  {\bibinfo {volume} {2013}},\ \bibinfo {pages} {033} (\bibinfo {year}
  {2013})},\ \Eprint {http://arxiv.org/abs/1307.6090} {arXiv:1307.6090}
  \BibitemShut {NoStop}%
\bibitem [{\citenamefont {{Nagy}}\ \emph {et~al.}(2017)\citenamefont {{Nagy}},
  \citenamefont {{Ade}}, \citenamefont {{Amiri}}, \citenamefont {{Benton}},
  \citenamefont {{Bergman}}, \citenamefont {{Bihary}}, \citenamefont {{Bock}},
  \citenamefont {{Bond}}, \citenamefont {{Bryan}}, \citenamefont {{Chiang}},
  \citenamefont {{Contaldi}}, \citenamefont {{Dor{\'e}}}, \citenamefont
  {{Duivenvoorden}}, \citenamefont {{Eriksen}}, \citenamefont {{Farhang}},
  \citenamefont {{Filippini}}, \citenamefont {{Fissel}}, \citenamefont
  {{Fraisse}}, \citenamefont {{Freese}}, \citenamefont {{Galloway}},
  \citenamefont {{Gambrel}}, \citenamefont {{Gandilo}}, \citenamefont
  {{Ganga}}, \citenamefont {{Gudmundsson}}, \citenamefont {{Halpern}},
  \citenamefont {{Hartley}}, \citenamefont {{Hasselfield}}, \citenamefont
  {{Hilton}}, \citenamefont {{Holmes}}, \citenamefont {{Hristov}},
  \citenamefont {{Huang}}, \citenamefont {{Irwin}}, \citenamefont {{Jones}},
  \citenamefont {{Kuo}}, \citenamefont {{Kermish}}, \citenamefont {{Li}},
  \citenamefont {{Mason}}, \citenamefont {{Megerian}}, \citenamefont
  {{Moncelsi}}, \citenamefont {{Morford}}, \citenamefont {{Netterfield}},
  \citenamefont {{Nolta}}, \citenamefont {{O'Brient}}, \citenamefont
  {{Osherson}}, \citenamefont {{Padilla}}, \citenamefont {{Racine}},
  \citenamefont {{Rahlin}}, \citenamefont {{Reintsema}}, \citenamefont
  {{Ruhl}}, \citenamefont {{Runyan}}, \citenamefont {{Ruud}}, \citenamefont
  {{Shariff}}, \citenamefont {{Shaw}}, \citenamefont {{Shiu}}, \citenamefont
  {{Soler}}, \citenamefont {{Song}}, \citenamefont {{Trangsrud}}, \citenamefont
  {{Tucker}}, \citenamefont {{Tucker}}, \citenamefont {{Turner}}, \citenamefont
  {{van der List}}, \citenamefont {{Weber}}, \citenamefont {{Wehus}},
  \citenamefont {{Wiebe}},\ and\ \citenamefont
  {{Young}}}]{2017ApJ...844..151N}%
  \BibitemOpen
  \bibfield  {author} {\bibinfo {author} {\bibfnamefont {J.~M.}\ \bibnamefont
  {{Nagy}}}, \bibinfo {author} {\bibfnamefont {P.~A.~R.}\ \bibnamefont
  {{Ade}}}, \bibinfo {author} {\bibfnamefont {M.}~\bibnamefont {{Amiri}}},
  \bibinfo {author} {\bibfnamefont {S.~J.}\ \bibnamefont {{Benton}}}, \bibinfo
  {author} {\bibfnamefont {A.~S.}\ \bibnamefont {{Bergman}}}, \bibinfo {author}
  {\bibfnamefont {R.}~\bibnamefont {{Bihary}}}, \bibinfo {author}
  {\bibfnamefont {J.~J.}\ \bibnamefont {{Bock}}}, \bibinfo {author}
  {\bibfnamefont {J.~R.}\ \bibnamefont {{Bond}}}, \bibinfo {author}
  {\bibfnamefont {S.~A.}\ \bibnamefont {{Bryan}}}, \bibinfo {author}
  {\bibfnamefont {H.~C.}\ \bibnamefont {{Chiang}}}, \bibinfo {author}
  {\bibfnamefont {C.~R.}\ \bibnamefont {{Contaldi}}}, \bibinfo {author}
  {\bibfnamefont {O.}~\bibnamefont {{Dor{\'e}}}}, \bibinfo {author}
  {\bibfnamefont {A.~J.}\ \bibnamefont {{Duivenvoorden}}}, \bibinfo {author}
  {\bibfnamefont {H.~K.}\ \bibnamefont {{Eriksen}}}, \bibinfo {author}
  {\bibfnamefont {M.}~\bibnamefont {{Farhang}}}, \bibinfo {author}
  {\bibfnamefont {J.~P.}\ \bibnamefont {{Filippini}}}, \bibinfo {author}
  {\bibfnamefont {L.~M.}\ \bibnamefont {{Fissel}}}, \bibinfo {author}
  {\bibfnamefont {A.~A.}\ \bibnamefont {{Fraisse}}}, \bibinfo {author}
  {\bibfnamefont {K.}~\bibnamefont {{Freese}}}, \bibinfo {author}
  {\bibfnamefont {M.}~\bibnamefont {{Galloway}}}, \bibinfo {author}
  {\bibfnamefont {A.~E.}\ \bibnamefont {{Gambrel}}}, \bibinfo {author}
  {\bibfnamefont {N.~N.}\ \bibnamefont {{Gandilo}}}, \bibinfo {author}
  {\bibfnamefont {K.}~\bibnamefont {{Ganga}}}, \bibinfo {author} {\bibfnamefont
  {J.~E.}\ \bibnamefont {{Gudmundsson}}}, \bibinfo {author} {\bibfnamefont
  {M.}~\bibnamefont {{Halpern}}}, \bibinfo {author} {\bibfnamefont
  {J.}~\bibnamefont {{Hartley}}}, \bibinfo {author} {\bibfnamefont
  {M.}~\bibnamefont {{Hasselfield}}}, \bibinfo {author} {\bibfnamefont
  {G.}~\bibnamefont {{Hilton}}}, \bibinfo {author} {\bibfnamefont
  {W.}~\bibnamefont {{Holmes}}}, \bibinfo {author} {\bibfnamefont {V.~V.}\
  \bibnamefont {{Hristov}}}, \bibinfo {author} {\bibfnamefont {Z.}~\bibnamefont
  {{Huang}}}, \bibinfo {author} {\bibfnamefont {K.~D.}\ \bibnamefont
  {{Irwin}}}, \bibinfo {author} {\bibfnamefont {W.~C.}\ \bibnamefont
  {{Jones}}}, \bibinfo {author} {\bibfnamefont {C.~L.}\ \bibnamefont {{Kuo}}},
  \bibinfo {author} {\bibfnamefont {Z.~D.}\ \bibnamefont {{Kermish}}}, \bibinfo
  {author} {\bibfnamefont {S.}~\bibnamefont {{Li}}}, \bibinfo {author}
  {\bibfnamefont {P.~V.}\ \bibnamefont {{Mason}}}, \bibinfo {author}
  {\bibfnamefont {K.}~\bibnamefont {{Megerian}}}, \bibinfo {author}
  {\bibfnamefont {L.}~\bibnamefont {{Moncelsi}}}, \bibinfo {author}
  {\bibfnamefont {T.~A.}\ \bibnamefont {{Morford}}}, \bibinfo {author}
  {\bibfnamefont {C.~B.}\ \bibnamefont {{Netterfield}}}, \bibinfo {author}
  {\bibfnamefont {M.}~\bibnamefont {{Nolta}}}, \bibinfo {author} {\bibfnamefont
  {R.}~\bibnamefont {{O'Brient}}}, \bibinfo {author} {\bibfnamefont
  {B.}~\bibnamefont {{Osherson}}}, \bibinfo {author} {\bibfnamefont {I.~L.}\
  \bibnamefont {{Padilla}}}, \bibinfo {author} {\bibfnamefont {B.}~\bibnamefont
  {{Racine}}}, \bibinfo {author} {\bibfnamefont {A.~S.}\ \bibnamefont
  {{Rahlin}}}, \bibinfo {author} {\bibfnamefont {C.}~\bibnamefont
  {{Reintsema}}}, \bibinfo {author} {\bibfnamefont {J.~E.}\ \bibnamefont
  {{Ruhl}}}, \bibinfo {author} {\bibfnamefont {M.~C.}\ \bibnamefont
  {{Runyan}}}, \bibinfo {author} {\bibfnamefont {T.~M.}\ \bibnamefont
  {{Ruud}}}, \bibinfo {author} {\bibfnamefont {J.~A.}\ \bibnamefont
  {{Shariff}}}, \bibinfo {author} {\bibfnamefont {E.~C.}\ \bibnamefont
  {{Shaw}}}, \bibinfo {author} {\bibfnamefont {C.}~\bibnamefont {{Shiu}}},
  \bibinfo {author} {\bibfnamefont {J.~D.}\ \bibnamefont {{Soler}}}, \bibinfo
  {author} {\bibfnamefont {X.}~\bibnamefont {{Song}}}, \bibinfo {author}
  {\bibfnamefont {A.}~\bibnamefont {{Trangsrud}}}, \bibinfo {author}
  {\bibfnamefont {C.}~\bibnamefont {{Tucker}}}, \bibinfo {author}
  {\bibfnamefont {R.~S.}\ \bibnamefont {{Tucker}}}, \bibinfo {author}
  {\bibfnamefont {A.~D.}\ \bibnamefont {{Turner}}}, \bibinfo {author}
  {\bibfnamefont {J.~F.}\ \bibnamefont {{van der List}}}, \bibinfo {author}
  {\bibfnamefont {A.~C.}\ \bibnamefont {{Weber}}}, \bibinfo {author}
  {\bibfnamefont {I.~K.}\ \bibnamefont {{Wehus}}}, \bibinfo {author}
  {\bibfnamefont {D.~V.}\ \bibnamefont {{Wiebe}}}, \ and\ \bibinfo {author}
  {\bibfnamefont {E.~Y.}\ \bibnamefont {{Young}}},\ }\href {\doibase
  10.3847/1538-4357/aa7cfd} {\bibfield  {journal} {\bibinfo  {journal} {\apj}\
  }\textbf {\bibinfo {volume} {844}},\ \bibinfo {eid} {151} (\bibinfo {year}
  {2017})},\ \Eprint {http://arxiv.org/abs/1704.00215} {arXiv:1704.00215
  [astro-ph.CO]} \BibitemShut {NoStop}%
\bibitem [{\citenamefont {{Padilla}}\ \emph {et~al.}(2020)\citenamefont
  {{Padilla}}, \citenamefont {{Eimer}}, \citenamefont {{Li}}, \citenamefont
  {{Addison}}, \citenamefont {{Ali}}, \citenamefont {{Appel}}, \citenamefont
  {{Bennett}}, \citenamefont {{Bustos}}, \citenamefont {{Brewer}},
  \citenamefont {{Chan}}, \citenamefont {{Chuss}}, \citenamefont {{Cleary}},
  \citenamefont {{Couto}}, \citenamefont {{Dahal}}, \citenamefont {{Denis}},
  \citenamefont {{D{\"u}nner}}, \citenamefont {{Essinger-Hileman}},
  \citenamefont {{Flux{\`a}}}, \citenamefont {{Gothe}}, \citenamefont
  {{Harrington}}, \citenamefont {{Iuliano}}, \citenamefont {{Karakla}},
  \citenamefont {{Marriage}}, \citenamefont {{Miller}}, \citenamefont
  {{N{\'u}{\~n}ez}}, \citenamefont {{Parker}}, \citenamefont {{Petroff}},
  \citenamefont {{Pradenas}}, \citenamefont {{Reeves}}, \citenamefont
  {{Rostem}}, \citenamefont {{Stevens}}, \citenamefont {{N{\'u}{\~n}ez}},
  \citenamefont {{Watts}}, \citenamefont {{Weiland}}, \citenamefont
  {{Wollack}}, \citenamefont {{Xu}},\ and\ \citenamefont {{CLASS
  Collaboration}}}]{2020ApJ...889..105P}%
  \BibitemOpen
  \bibfield  {author} {\bibinfo {author} {\bibfnamefont {I.~L.}\ \bibnamefont
  {{Padilla}}}, \bibinfo {author} {\bibfnamefont {J.~R.}\ \bibnamefont
  {{Eimer}}}, \bibinfo {author} {\bibfnamefont {Y.}~\bibnamefont {{Li}}},
  \bibinfo {author} {\bibfnamefont {G.~E.}\ \bibnamefont {{Addison}}}, \bibinfo
  {author} {\bibfnamefont {A.}~\bibnamefont {{Ali}}}, \bibinfo {author}
  {\bibfnamefont {J.~W.}\ \bibnamefont {{Appel}}}, \bibinfo {author}
  {\bibfnamefont {C.~L.}\ \bibnamefont {{Bennett}}}, \bibinfo {author}
  {\bibfnamefont {R.}~\bibnamefont {{Bustos}}}, \bibinfo {author}
  {\bibfnamefont {M.~K.}\ \bibnamefont {{Brewer}}}, \bibinfo {author}
  {\bibfnamefont {M.}~\bibnamefont {{Chan}}}, \bibinfo {author} {\bibfnamefont
  {D.~T.}\ \bibnamefont {{Chuss}}}, \bibinfo {author} {\bibfnamefont
  {J.}~\bibnamefont {{Cleary}}}, \bibinfo {author} {\bibfnamefont
  {J.}~\bibnamefont {{Couto}}}, \bibinfo {author} {\bibfnamefont
  {S.}~\bibnamefont {{Dahal}}}, \bibinfo {author} {\bibfnamefont
  {K.}~\bibnamefont {{Denis}}}, \bibinfo {author} {\bibfnamefont
  {R.}~\bibnamefont {{D{\"u}nner}}}, \bibinfo {author} {\bibfnamefont
  {T.}~\bibnamefont {{Essinger-Hileman}}}, \bibinfo {author} {\bibfnamefont
  {P.}~\bibnamefont {{Flux{\`a}}}}, \bibinfo {author} {\bibfnamefont
  {D.}~\bibnamefont {{Gothe}}}, \bibinfo {author} {\bibfnamefont
  {K.}~\bibnamefont {{Harrington}}}, \bibinfo {author} {\bibfnamefont
  {J.}~\bibnamefont {{Iuliano}}}, \bibinfo {author} {\bibfnamefont
  {J.}~\bibnamefont {{Karakla}}}, \bibinfo {author} {\bibfnamefont {T.~A.}\
  \bibnamefont {{Marriage}}}, \bibinfo {author} {\bibfnamefont {N.~J.}\
  \bibnamefont {{Miller}}}, \bibinfo {author} {\bibfnamefont {C.}~\bibnamefont
  {{N{\'u}{\~n}ez}}}, \bibinfo {author} {\bibfnamefont {L.}~\bibnamefont
  {{Parker}}}, \bibinfo {author} {\bibfnamefont {M.}~\bibnamefont {{Petroff}}},
  \bibinfo {author} {\bibfnamefont {B.}~\bibnamefont {{Pradenas}}}, \bibinfo
  {author} {\bibfnamefont {R.}~\bibnamefont {{Reeves}}}, \bibinfo {author}
  {\bibfnamefont {K.}~\bibnamefont {{Rostem}}}, \bibinfo {author}
  {\bibfnamefont {D.~T.}\ \bibnamefont {{Stevens}}}, \bibinfo {author}
  {\bibfnamefont {D.~A.~V.}\ \bibnamefont {{N{\'u}{\~n}ez}}}, \bibinfo {author}
  {\bibfnamefont {D.~J.}\ \bibnamefont {{Watts}}}, \bibinfo {author}
  {\bibfnamefont {J.~L.}\ \bibnamefont {{Weiland}}}, \bibinfo {author}
  {\bibfnamefont {E.~J.}\ \bibnamefont {{Wollack}}}, \bibinfo {author}
  {\bibfnamefont {Z.}~\bibnamefont {{Xu}}}, \ and\ \bibinfo {author}
  {\bibnamefont {{CLASS Collaboration}}},\ }\href {\doibase
  10.3847/1538-4357/ab61f8} {\bibfield  {journal} {\bibinfo  {journal} {\apj}\
  }\textbf {\bibinfo {volume} {889}},\ \bibinfo {eid} {105} (\bibinfo {year}
  {2020})},\ \Eprint {http://arxiv.org/abs/1911.00391} {arXiv:1911.00391
  [astro-ph.CO]} \BibitemShut {NoStop}%
\bibitem [{\citenamefont {Alexander}\ \emph {et~al.}(2020)\citenamefont
  {Alexander}, \citenamefont {McDonough}, \citenamefont {Pullen},\ and\
  \citenamefont {Shapiro}}]{alexander2020physics}%
  \BibitemOpen
  \bibfield  {author} {\bibinfo {author} {\bibfnamefont {S.}~\bibnamefont
  {Alexander}}, \bibinfo {author} {\bibfnamefont {E.}~\bibnamefont
  {McDonough}}, \bibinfo {author} {\bibfnamefont {A.}~\bibnamefont {Pullen}}, \
  and\ \bibinfo {author} {\bibfnamefont {B.}~\bibnamefont {Shapiro}},\ }\href
  {\doibase 10.1088/1475-7516/2020/01/032} {\bibfield  {journal} {\bibinfo
  {journal} {Journal of Cosmology and Astroparticle Physics}\ }\textbf
  {\bibinfo {volume} {2020}},\ \bibinfo {pages} {032} (\bibinfo {year}
  {2020})},\ \Eprint {http://arxiv.org/abs/1911.01418} {arXiv:1911.01418}
  \BibitemShut {NoStop}%
\bibitem [{\citenamefont {Giovannini}(2010)}]{giovannini2010circular}%
  \BibitemOpen
  \bibfield  {author} {\bibinfo {author} {\bibfnamefont {M.}~\bibnamefont
  {Giovannini}},\ }\href@noop {} {\bibfield  {journal} {\bibinfo  {journal}
  {Physical Review D}\ }\textbf {\bibinfo {volume} {81}},\ \bibinfo {pages}
  {023003} (\bibinfo {year} {2010})},\ \Eprint {http://arxiv.org/abs/0909.4699}
  {arXiv:0909.4699} \BibitemShut {NoStop}%
\bibitem [{\citenamefont {Giovannini}(2009)}]{giovannini2009vmode}%
  \BibitemOpen
  \bibfield  {author} {\bibinfo {author} {\bibfnamefont {M.}~\bibnamefont
  {Giovannini}},\ }\href@noop {} {\bibfield  {journal} {\bibinfo  {journal}
  {Physical Review D}\ }\textbf {\bibinfo {volume} {80}},\ \bibinfo {pages}
  {123013} (\bibinfo {year} {2009})},\ \Eprint {http://arxiv.org/abs/0909.3629}
  {arXiv:0909.3629} \BibitemShut {NoStop}%
\bibitem [{\citenamefont {Zarei}\ \emph {et~al.}(2010)\citenamefont {Zarei},
  \citenamefont {Bavarsad}, \citenamefont {Haghighat}, \citenamefont
  {Mohammadi}, \citenamefont {Motie},\ and\ \citenamefont
  {Rezaei}}]{zarei2010generation}%
  \BibitemOpen
  \bibfield  {author} {\bibinfo {author} {\bibfnamefont {M.}~\bibnamefont
  {Zarei}}, \bibinfo {author} {\bibfnamefont {E.}~\bibnamefont {Bavarsad}},
  \bibinfo {author} {\bibfnamefont {M.}~\bibnamefont {Haghighat}}, \bibinfo
  {author} {\bibfnamefont {R.}~\bibnamefont {Mohammadi}}, \bibinfo {author}
  {\bibfnamefont {I.}~\bibnamefont {Motie}}, \ and\ \bibinfo {author}
  {\bibfnamefont {Z.}~\bibnamefont {Rezaei}},\ }\href@noop {} {\bibfield
  {journal} {\bibinfo  {journal} {Physical Review D}\ }\textbf {\bibinfo
  {volume} {81}},\ \bibinfo {pages} {084035} (\bibinfo {year}
  {2010})}\BibitemShut {NoStop}%
\bibitem [{\citenamefont {Tizchang}\ \emph {et~al.}(2016)\citenamefont
  {Tizchang}, \citenamefont {Batebi}, \citenamefont {Haghighat},\ and\
  \citenamefont {Mohammadi}}]{tizchang2016noncommutative}%
  \BibitemOpen
  \bibfield  {author} {\bibinfo {author} {\bibfnamefont {S.}~\bibnamefont
  {Tizchang}}, \bibinfo {author} {\bibfnamefont {S.}~\bibnamefont {Batebi}},
  \bibinfo {author} {\bibfnamefont {M.}~\bibnamefont {Haghighat}}, \ and\
  \bibinfo {author} {\bibfnamefont {R.}~\bibnamefont {Mohammadi}},\ }\href
  {\doibase 10.1140/epjc/s10052-016-4312-5} {\bibfield  {journal} {\bibinfo
  {journal} {European Physical Journal C}\ }\textbf {\bibinfo {volume} {76}},\
  \bibinfo {pages} {478} (\bibinfo {year} {2016})},\ \Eprint
  {http://arxiv.org/abs/1605.09045} {arXiv:1605.09045} \BibitemShut {NoStop}%
\bibitem [{\citenamefont {{Alexander}}\ \emph {et~al.}(2009)\citenamefont
  {{Alexander}}, \citenamefont {{Ochoa}},\ and\ \citenamefont
  {{Kosowsky}}}]{2009PhRvD..79f3524A}%
  \BibitemOpen
  \bibfield  {author} {\bibinfo {author} {\bibfnamefont {S.}~\bibnamefont
  {{Alexander}}}, \bibinfo {author} {\bibfnamefont {J.}~\bibnamefont
  {{Ochoa}}}, \ and\ \bibinfo {author} {\bibfnamefont {A.}~\bibnamefont
  {{Kosowsky}}},\ }\href {\doibase 10.1103/PhysRevD.79.063524} {\bibfield
  {journal} {\bibinfo  {journal} {Physical Review D}\ }\textbf {\bibinfo
  {volume} {79}},\ \bibinfo {eid} {063524} (\bibinfo {year} {2009})},\ \Eprint
  {http://arxiv.org/abs/0810.2355} {arXiv:0810.2355 [astro-ph]} \BibitemShut
  {NoStop}%
\bibitem [{\citenamefont {Cooray}\ \emph {et~al.}(2003)\citenamefont {Cooray},
  \citenamefont {Melchiorri},\ and\ \citenamefont
  {Silk}}]{cooray2003circularly}%
  \BibitemOpen
  \bibfield  {author} {\bibinfo {author} {\bibfnamefont {A.}~\bibnamefont
  {Cooray}}, \bibinfo {author} {\bibfnamefont {A.}~\bibnamefont {Melchiorri}},
  \ and\ \bibinfo {author} {\bibfnamefont {J.}~\bibnamefont {Silk}},\ }\href
  {\doibase 10.1016/S0370-2693(02)03291-4} {\bibfield  {journal} {\bibinfo
  {journal} {Physics Letters B}\ }\textbf {\bibinfo {volume} {554}},\ \bibinfo
  {pages} {1} (\bibinfo {year} {2003})},\ \Eprint
  {http://arxiv.org/abs/astro-ph/0205214} {arXiv:astro-ph/0205214} \BibitemShut
  {NoStop}%
\bibitem [{\citenamefont {Montero-Camacho}\ and\ \citenamefont
  {Hirata}(2018)}]{montero2018exploring}%
  \BibitemOpen
  \bibfield  {author} {\bibinfo {author} {\bibfnamefont {P.}~\bibnamefont
  {Montero-Camacho}}\ and\ \bibinfo {author} {\bibfnamefont {C.~M.}\
  \bibnamefont {Hirata}},\ }\href@noop {} {\bibfield  {journal} {\bibinfo
  {journal} {Journal of Cosmology and Astroparticle Physics}\ }\textbf
  {\bibinfo {volume} {2018}},\ \bibinfo {pages} {040} (\bibinfo {year}
  {2018})}\BibitemShut {NoStop}%
\bibitem [{\citenamefont {Agarwal}\ \emph {et~al.}(2008)\citenamefont
  {Agarwal}, \citenamefont {Jain}, \citenamefont {McKay},\ and\ \citenamefont
  {Ralston}}]{agarwal2008signatures}%
  \BibitemOpen
  \bibfield  {author} {\bibinfo {author} {\bibfnamefont {N.}~\bibnamefont
  {Agarwal}}, \bibinfo {author} {\bibfnamefont {P.}~\bibnamefont {Jain}},
  \bibinfo {author} {\bibfnamefont {D.~W.}\ \bibnamefont {McKay}}, \ and\
  \bibinfo {author} {\bibfnamefont {J.~P.}\ \bibnamefont {Ralston}},\ }\href
  {\doibase 10.1103/PhysRevD.78.085028} {\bibfield  {journal} {\bibinfo
  {journal} {Physical Review D}\ }\textbf {\bibinfo {volume} {78}},\ \bibinfo
  {pages} {085028} (\bibinfo {year} {2008})},\ \Eprint
  {http://arxiv.org/abs/0807.4587} {arXiv:0807.4587} \BibitemShut {NoStop}%
\bibitem [{\citenamefont {Finelli}\ and\ \citenamefont
  {Galaverni}(2009)}]{finelli2009rotation}%
  \BibitemOpen
  \bibfield  {author} {\bibinfo {author} {\bibfnamefont {F.}~\bibnamefont
  {Finelli}}\ and\ \bibinfo {author} {\bibfnamefont {M.}~\bibnamefont
  {Galaverni}},\ }\href {\doibase 10.1103/PhysRevD.79.063002} {\bibfield
  {journal} {\bibinfo  {journal} {Physical Review D}\ }\textbf {\bibinfo
  {volume} {79}},\ \bibinfo {pages} {063002} (\bibinfo {year} {2009})},\
  \Eprint {http://arxiv.org/abs/0802.4210} {arXiv:0802.4210} \BibitemShut
  {NoStop}%
\bibitem [{\citenamefont {Mohammadi}(2014)}]{mohammadi2014evidence}%
  \BibitemOpen
  \bibfield  {author} {\bibinfo {author} {\bibfnamefont {R.}~\bibnamefont
  {Mohammadi}},\ }\href {\doibase 10.1140/epjc/s10052-014-3102-2} {\bibfield
  {journal} {\bibinfo  {journal} {The European Physical Journal C}\ }\textbf
  {\bibinfo {volume} {74}},\ \bibinfo {pages} {3102} (\bibinfo {year}
  {2014})},\ \Eprint {http://arxiv.org/abs/1312.2199} {arXiv:1312.2199}
  \BibitemShut {NoStop}%
\bibitem [{\citenamefont {Bartolo}\ \emph {et~al.}(2019)\citenamefont
  {Bartolo}, \citenamefont {Hoseinpour}, \citenamefont {Matarrese},
  \citenamefont {Orlando},\ and\ \citenamefont {Zarei}}]{bartolo2019cmb}%
  \BibitemOpen
  \bibfield  {author} {\bibinfo {author} {\bibfnamefont {N.}~\bibnamefont
  {Bartolo}}, \bibinfo {author} {\bibfnamefont {A.}~\bibnamefont {Hoseinpour}},
  \bibinfo {author} {\bibfnamefont {S.}~\bibnamefont {Matarrese}}, \bibinfo
  {author} {\bibfnamefont {G.}~\bibnamefont {Orlando}}, \ and\ \bibinfo
  {author} {\bibfnamefont {M.}~\bibnamefont {Zarei}},\ }\href {\doibase
  10.1103/PhysRevD.100.043516} {\bibfield  {journal} {\bibinfo  {journal}
  {Physical Review D}\ }\textbf {\bibinfo {volume} {100}},\ \bibinfo {pages}
  {043516} (\bibinfo {year} {2019})},\ \Eprint
  {http://arxiv.org/abs/1903.04578} {arXiv:1903.04578} \BibitemShut {NoStop}%
\bibitem [{\citenamefont {Haghighat}\ \emph {et~al.}(2020)\citenamefont
  {Haghighat}, \citenamefont {Mahmoudi}, \citenamefont {Mohammadi},
  \citenamefont {Tizchang},\ and\ \citenamefont {Xue}}]{haghighat2020sterile}%
  \BibitemOpen
  \bibfield  {author} {\bibinfo {author} {\bibfnamefont {M.}~\bibnamefont
  {Haghighat}}, \bibinfo {author} {\bibfnamefont {S.}~\bibnamefont {Mahmoudi}},
  \bibinfo {author} {\bibfnamefont {R.}~\bibnamefont {Mohammadi}}, \bibinfo
  {author} {\bibfnamefont {S.}~\bibnamefont {Tizchang}}, \ and\ \bibinfo
  {author} {\bibfnamefont {S.-S.}\ \bibnamefont {Xue}},\ }\href {\doibase
  10.1103/PhysRevD.101.123016} {\bibfield  {journal} {\bibinfo  {journal}
  {Physical Review D}\ }\textbf {\bibinfo {volume} {101}},\ \bibinfo {pages}
  {123016} (\bibinfo {year} {2020})},\ \Eprint
  {http://arxiv.org/abs/1909.03883} {arXiv:1909.03883} \BibitemShut {NoStop}%
\bibitem [{\citenamefont {Hoseinpour}\ \emph {et~al.}(2020)\citenamefont
  {Hoseinpour}, \citenamefont {Zarei}, \citenamefont {Orlando}, \citenamefont
  {Bartolo},\ and\ \citenamefont {Matarrese}}]{hoseinpour2020cmb}%
  \BibitemOpen
  \bibfield  {author} {\bibinfo {author} {\bibfnamefont {A.}~\bibnamefont
  {Hoseinpour}}, \bibinfo {author} {\bibfnamefont {M.}~\bibnamefont {Zarei}},
  \bibinfo {author} {\bibfnamefont {G.}~\bibnamefont {Orlando}}, \bibinfo
  {author} {\bibfnamefont {N.}~\bibnamefont {Bartolo}}, \ and\ \bibinfo
  {author} {\bibfnamefont {S.}~\bibnamefont {Matarrese}},\ }\href {\doibase
  10.1103/PhysRevD.102.063501} {\bibfield  {journal} {\bibinfo  {journal}
  {Physical Review D}\ }\textbf {\bibinfo {volume} {102}},\ \bibinfo {pages}
  {063501} (\bibinfo {year} {2020})},\ \Eprint
  {http://arxiv.org/abs/2006.14418} {arXiv:2006.14418} \BibitemShut {NoStop}%
\bibitem [{\citenamefont {Sawyer}(2015)}]{sawyer2015photonphoton}%
  \BibitemOpen
  \bibfield  {author} {\bibinfo {author} {\bibfnamefont {R.~F.}\ \bibnamefont
  {Sawyer}},\ }\href {\doibase 10.1103/PhysRevD.91.021301} {\bibfield
  {journal} {\bibinfo  {journal} {Physical Review D}\ }\textbf {\bibinfo
  {volume} {91}},\ \bibinfo {pages} {021301} (\bibinfo {year} {2015})},\
  \Eprint {http://arxiv.org/abs/1408.5434} {arXiv:1408.5434} \BibitemShut
  {NoStop}%
\bibitem [{\citenamefont {Inomata}\ and\ \citenamefont
  {Kamionkowski}(2019)}]{inomata2019circular}%
  \BibitemOpen
  \bibfield  {author} {\bibinfo {author} {\bibfnamefont {K.}~\bibnamefont
  {Inomata}}\ and\ \bibinfo {author} {\bibfnamefont {M.}~\bibnamefont
  {Kamionkowski}},\ }\href {\doibase 10.1103/PhysRevD.99.043501} {\bibfield
  {journal} {\bibinfo  {journal} {Physical Review D}\ }\textbf {\bibinfo
  {volume} {99}},\ \bibinfo {pages} {043501} (\bibinfo {year} {2019})},\
  \Eprint {http://arxiv.org/abs/1811.04957} {arXiv:1811.04957} \BibitemShut
  {NoStop}%
\bibitem [{\citenamefont {{Holdom}}(1986)}]{1986PhLB..166..196H}%
  \BibitemOpen
  \bibfield  {author} {\bibinfo {author} {\bibfnamefont {B.}~\bibnamefont
  {{Holdom}}},\ }\href {\doibase 10.1016/0370-2693(86)91377-8} {\bibfield
  {journal} {\bibinfo  {journal} {Physics Letters B}\ }\textbf {\bibinfo
  {volume} {166}},\ \bibinfo {pages} {196} (\bibinfo {year}
  {1986})}\BibitemShut {NoStop}%
\bibitem [{\citenamefont {Fabbrichesi}\ \emph {et~al.}(2021)\citenamefont
  {Fabbrichesi}, \citenamefont {Gabrielli},\ and\ \citenamefont
  {Lanfranchi}}]{fabbrichesi2020dark}%
  \BibitemOpen
  \bibfield  {author} {\bibinfo {author} {\bibfnamefont {M.}~\bibnamefont
  {Fabbrichesi}}, \bibinfo {author} {\bibfnamefont {E.}~\bibnamefont
  {Gabrielli}}, \ and\ \bibinfo {author} {\bibfnamefont {G.}~\bibnamefont
  {Lanfranchi}},\ }\href {\doibase 10.1007/978-3-030-62519-1} {\emph {\bibinfo
  {title} {The Physics of the Dark Photon: A Primer}}},\ SpringerBriefs in
  Physics\ (\bibinfo  {publisher} {Springer},\ \bibinfo {address} {Cham},\
  \bibinfo {year} {2021})\ \Eprint {http://arxiv.org/abs/2005.01515}
  {arXiv:2005.01515} \BibitemShut {NoStop}%
\bibitem [{\citenamefont {Caputo}\ \emph {et~al.}(2021)\citenamefont {Caputo},
  \citenamefont {Millar}, \citenamefont {O’Hare},\ and\ \citenamefont
  {Vitagliano}}]{caputo2021dark}%
  \BibitemOpen
  \bibfield  {author} {\bibinfo {author} {\bibfnamefont {A.}~\bibnamefont
  {Caputo}}, \bibinfo {author} {\bibfnamefont {A.~J.}\ \bibnamefont {Millar}},
  \bibinfo {author} {\bibfnamefont {C.~A.}\ \bibnamefont {O’Hare}}, \ and\
  \bibinfo {author} {\bibfnamefont {E.}~\bibnamefont {Vitagliano}},\
  }\href@noop {} {\bibfield  {journal} {\bibinfo  {journal} {Physical Review
  D}\ }\textbf {\bibinfo {volume} {104}},\ \bibinfo {pages} {095029} (\bibinfo
  {year} {2021})}\BibitemShut {NoStop}%
\bibitem [{\citenamefont {Caputo}\ \emph {et~al.}(2026)\citenamefont {Caputo},
  \citenamefont {Park},\ and\ \citenamefont {Yun}}]{caputo2025heavy}%
  \BibitemOpen
  \bibfield  {author} {\bibinfo {author} {\bibfnamefont {A.}~\bibnamefont
  {Caputo}}, \bibinfo {author} {\bibfnamefont {J.}~\bibnamefont {Park}}, \ and\
  \bibinfo {author} {\bibfnamefont {S.}~\bibnamefont {Yun}},\ }\href {\doibase
  10.1103/m6zs-jxxp} {\bibfield  {journal} {\bibinfo  {journal} {Physical
  Review D}\ }\textbf {\bibinfo {volume} {113}},\ \bibinfo {pages} {075001}
  (\bibinfo {year} {2026})},\ \Eprint {http://arxiv.org/abs/2511.15785}
  {arXiv:2511.15785} \BibitemShut {NoStop}%
\bibitem [{\citenamefont {{Heisenberg}}\ and\ \citenamefont
  {{Euler}}(1936)}]{2006physics...5038H}%
  \BibitemOpen
  \bibfield  {author} {\bibinfo {author} {\bibfnamefont {W.}~\bibnamefont
  {{Heisenberg}}}\ and\ \bibinfo {author} {\bibfnamefont {H.}~\bibnamefont
  {{Euler}}},\ }\href {\doibase 10.1007/BF01343663} {\bibfield  {journal}
  {\bibinfo  {journal} {Zeitschrift f{\"u}r Physik}\ }\textbf {\bibinfo
  {volume} {98}},\ \bibinfo {pages} {714} (\bibinfo {year} {1936})},\ \Eprint
  {http://arxiv.org/abs/physics/0605038} {arXiv:physics/0605038
  [physics.hist-ph]} \BibitemShut {NoStop}%
\bibitem [{\citenamefont {Georgi}(1993)}]{georgi1993effective}%
  \BibitemOpen
  \bibfield  {author} {\bibinfo {author} {\bibfnamefont {H.}~\bibnamefont
  {Georgi}},\ }\href {\doibase 10.1146/annurev.ns.43.120193.001233} {\bibfield
  {journal} {\bibinfo  {journal} {Annual Review of Nuclear and Particle
  Science}\ }\textbf {\bibinfo {volume} {43}},\ \bibinfo {pages} {209}
  (\bibinfo {year} {1993})}\BibitemShut {NoStop}%
\bibitem [{\citenamefont {Kosowsky}(2002)}]{kosowsky2001cmb}%
  \BibitemOpen
  \bibfield  {author} {\bibinfo {author} {\bibfnamefont {A.}~\bibnamefont
  {Kosowsky}},\ }in\ \href@noop {} {\emph {\bibinfo {booktitle} {Modern
  Cosmology}}},\ \bibinfo {editor} {edited by\ \bibinfo {editor} {\bibfnamefont
  {S.}~\bibnamefont {Bonometto}}, \bibinfo {editor} {\bibfnamefont
  {V.}~\bibnamefont {Gorini}}, \ and\ \bibinfo {editor} {\bibfnamefont
  {U.}~\bibnamefont {Moschella}}}\ (\bibinfo  {publisher} {IOP Publishing},\
  \bibinfo {year} {2002})\ p.\ \bibinfo {pages} {219},\ \Eprint
  {http://arxiv.org/abs/astro-ph/0102402} {arXiv:astro-ph/0102402} \BibitemShut
  {NoStop}%
\bibitem [{\citenamefont {Lin}\ and\ \citenamefont
  {Wandelt}(2006)}]{lin2006beginners}%
  \BibitemOpen
  \bibfield  {author} {\bibinfo {author} {\bibfnamefont {Y.-T.}\ \bibnamefont
  {Lin}}\ and\ \bibinfo {author} {\bibfnamefont {B.~D.}\ \bibnamefont
  {Wandelt}},\ }\href {\doibase 10.1016/j.astropartphys.2005.12.005} {\bibfield
   {journal} {\bibinfo  {journal} {Astroparticle Physics}\ }\textbf {\bibinfo
  {volume} {25}},\ \bibinfo {pages} {151} (\bibinfo {year} {2006})},\ \Eprint
  {http://arxiv.org/abs/astro-ph/0409734} {arXiv:astro-ph/0409734} \BibitemShut
  {NoStop}%
\bibitem [{\citenamefont {Schmidt}\ and\ \citenamefont
  {Jeong}(2012)}]{schmidt2012cosmic}%
  \BibitemOpen
  \bibfield  {author} {\bibinfo {author} {\bibfnamefont {F.}~\bibnamefont
  {Schmidt}}\ and\ \bibinfo {author} {\bibfnamefont {D.}~\bibnamefont
  {Jeong}},\ }\href {\doibase 10.1103/PhysRevD.86.083527} {\bibfield  {journal}
  {\bibinfo  {journal} {Physical Review D}\ }\textbf {\bibinfo {volume} {86}},\
  \bibinfo {pages} {083527} (\bibinfo {year} {2012})},\ \Eprint
  {http://arxiv.org/abs/1204.3625} {arXiv:1204.3625} \BibitemShut {NoStop}%
\bibitem [{\citenamefont {Wigner}(2012)}]{wigner2012group}%
  \BibitemOpen
  \bibfield  {author} {\bibinfo {author} {\bibfnamefont {E.~P.}\ \bibnamefont
  {Wigner}},\ }\href@noop {} {\emph {\bibinfo {title} {Group Theory and its
  Application to the Quantum Mechanics of Atomic Spectra}}}\ (\bibinfo
  {publisher} {Academic Press},\ \bibinfo {year} {2012})\BibitemShut {NoStop}%
\end{thebibliography}%

\end{document}